\documentclass{iopart}
\usepackage{iopams}  

\usepackage{epsfig}
\usepackage{graphicx}
\usepackage{subfigure}

\begin{document}

\title[]{Optimally chosen Nakajima-Zwanzig master equation for mean field approximation}

\author{Joshua Wilkie$^{1}$ and Yin Mei Wong$^2$}
\address{$^1$ Department of Chemistry, Simon Fraser University, Burnaby, British Columbia V5A 1S6, Canada\\
$^2$ Innovative Stochastic Algorithms, Surrey, British Columbia V3V 3N4, Canada}
\ead{wilkie@sfu.ca,yinwong01@shaw.ca}

\begin{abstract}

We define an ensemble of projection operators, each of which has an exact associated Nakajima-Zwanzig master equation for quantum open system evolution. A mean field approximation for the memory kernels is introduced that yields, for an optimally chosen projection operator, a completely determined inhomogeneous master equation. Previous proofs of positivity and equilibration are extended to these new inhomogeneous non-Markovian master equations. We study a nitrogen vacancy center in diamond interacting with $^{13}$C impurities to illustrate the method. 

\end{abstract}

\pacs{03.65.-w,05.30.-d,03.65.Yz}

\submitto{\JPA}
\maketitle

\section{Introduction}

Non-Markovian generalizations of the well known Kossakowski-Lindblad master equation\cite{dsg} have been a focus of many recent studies\cite{BV,BG,KR,KR2,KR3,Bud,WW3,WW1,W1,CK,CKP,BP2,AB,VB,CKR,LPB,BLP} in quantum open systems theory\cite{BP}. Such structures can preserve complete positivity\cite{BV,BG,KR,KR2,KR3,Bud} or positivity\cite{WW3,WW1,W1}. Efforts to quantify the non-Markovian character of such dynamics\cite{CKR,LPB,BLP} are also underway, as are explorations of relations between integrodifferential and time local forms\cite{CK,AB}. The interest stems partly from the fact that equilibration in a Markovian model cannot simultaneously be complete positivity preserving and correct in the long time limit except in certain exceptional cases\cite{THERM1,THERM2}. However, it is also true that non-Markovian effects are expected to play an important role in the low temperature condensed phase architectures\cite{BP} proposed for things like quantum computers\cite{QC}. 

Under certain conditions positivity preserving non-Markovian generalizations of the Kossakowski-Lindblad master equation\cite{dsg} may also be shown to correctly equilibrate\cite{WW3}. As with the Kossakowski-Lindblad master equation\cite{dsg} it is important to connect the parameters and memory functions of these abstract mathematical structures to data for specific physical systems. Semi-classical methods\cite{Wilk2} and mean field approaches\cite{Wilk} have been explored. Here we examine a modification of the mean field method to connect a specific positivity preserving master equation which correctly equilibrates\cite{WW3} to data for an electronic spin $S=1$ associated with a Nitrogen-Vacancy center in diamond interacting with the $I=1/2$ nuclear spins of $^{13}$C impurities. 

We start in Section 2 with a form of the exact Nakajima-Zwanzig master equation\cite{Zwan} (NZME) which is valid for cases in which the bath operator $\Lambda$, associated with the NZME projection operator
\begin{equation}
P={\cal I}_S\otimes \Lambda ~{\rm tr}_B\{\cdot\},\label{PO}
\end{equation}
commutes with the bath Hamiltonian. [${\cal I}_S$ is the identity operator in the system Hilbert space. Note that we will use ${\rm Tr}_B$ to denote a full trace over bath degrees of freedom and ${\rm tr}_B$ to denote the partial trace.] 

An ensemble of $\Lambda$ operators is defined in Section 3 wherein each $\Lambda$ is specified by setting the values of a set of scalar parameters. The NZME is exact for every parameter set. We then define a mean field approximation for this ensemble. Because the mean field master equations are approximate each will have a different accuracy. Most of the mean field master equations also possess an inhomogeneous term not previously considered in the abstract theory\cite{WW3}. With the exception of Ref. \cite{W1} previous studies of non-Markovian structures\cite{Bud,WW3,WW1,W1} did not consider inhomogeneous cases. Unfortunately the sufficient conditions discussed in Ref. \cite{W1} are inapplicable here.

In Section 4 we thus reexamine the issues of positivity and equilibration for this more general inhomogeneous structure. We find that it is sufficient for the memory functions to satisfy a set of constraints for positivity and equilibration to be guaranteed. Section 5 discusses the necessary basic structure of a model memory function that is consistent with the dictates of Section 4. In Section 6 we select the scalar parameters of the projection operator {\em and} those of the abstract theory such that the memory functions of the two different theories are optimally matched. Through this synthesis we are able to determine specific values of all parameters for the structure which preserves positivity and correctly equilibrates.

In Section 7 we discuss the results obtained by applying the synthetic theory to a simplified model of an NV center in diamond. The predictions of the synthetic master equation (SME) are compared to exact calculations obtained by well tested techniques\cite{TW}. Good accuracy is obtained at short and intermediate times but problems are seen in the long time limit. The diagonal elements are accurate within 5 \% but the off-diagonal elements, which in the exact case have persistent oscillations, are inaccurate. We discuss possible modifications of the abstract structures to correct this problem in Section 8.

\section{Nakajima-Zwanzig master equation}

Consider a simple composite system-bath with total Hamiltonian 
\begin{equation}
H_{tot}=H\otimes {\cal I}_B+S\otimes B+{\cal I}_S\otimes H_B.\label{HAM1}
\end{equation}
where $H$ and $H_B$ are the system and bath Hamiltonians, $S$ and $B$ are system and bath coupling operators, and $H$ and $S$ are in general non-commuting. Here ${\cal I}_B$ is the bath identity operator. We will employ units such that energies are measured in GHz and time is measured in ns. $S$ will be unit-less and $H$, $B$ and $H_B$ will have energy units.

We define a projection operator $P$ via (\ref{PO}) and choose the $\Lambda$ operator for the bath with the following properties:
\begin{eqnarray}
\Lambda^{\dag}&=&\Lambda\label{COND1}\\
{\rm Tr}_B\{\Lambda\}&=&1\\
\left[H_B,\Lambda \right]&=&0.\label{COND3}
\end{eqnarray}
It is important to note that $\Lambda\neq\rho_B$ in general where $\rho_B=e^{-H_B/k_BT}/Z_B$ is the canonical density operator and $Z_B={\rm Tr}_B\{e^{-H_B/k_BT}\}$. We also define a second projection operator via
\begin{equation}
Q={\cal I}_B-P
\end{equation}
so that $P+Q={\cal I}_B$, $P^2=P$ and $Q^2=Q$, and $PQ=QP=0$.

The total density operator $\rho_{tot}(t)=e^{-iH_{tot}t}\rho_{tot}(0)e^{iH_{tot}t}$ obeys the quantum Liouville equation
\begin{equation}
\frac{d\rho_{tot}(t)}{dt}=-iL_{tot}\rho_{tot}(t)
\end{equation}
where $L_{tot}=[H_{tot},\cdot ]$. It can then be shown that
\begin{eqnarray}
\frac{dP\rho_{tot}(t)}{dt}&=&-iPL_{tot}P\rho_{tot}(t)-iPL_{tot}Q\rho_{tot}(t)\label{EQ1}\\
\frac{dQ\rho_{tot}(t)}{dt}&=&-iQL_{tot}P\rho_{tot}(t)-iQL_{tot}Q\rho_{tot}(t)\label{EQ2}
\end{eqnarray}
and solving Eq. (\ref{EQ2}) for 
\begin{equation}
Q\rho_{tot}(t)=e^{-iQL_{tot}t}Q\rho_{tot}(0)-i\int_0^tdt'~e^{-iQL_{tot}(t-t')}QL_{tot}P\rho_{tot}(t')
\end{equation}
and substituting back into Eq. (\ref{EQ1}) gives
\begin{eqnarray}
\frac{dP\rho_{tot}(t)}{dt}&=&-iPL_{tot}e^{-iQL_{tot}t}Q\rho_{tot}(0)-iPL_{tot}P\rho_{tot}(t)\nonumber \\
&-&\int_0^tdt'~PL_{tot}Qe^{-iQL_{tot}(t-t')}QL_{tot}P\rho_{tot}(t').\label{ME1}
\end{eqnarray}

We choose an initial state of the form 
\begin{equation}
\rho_{tot}(0)=\rho(0)\otimes \rho_B
\end{equation}
with pure initial system state
\begin{equation}
\rho(0)=|\psi(0)\rangle\langle \psi(0)|.\label{INIT}
\end{equation}
Note that 
\begin{equation}
P\rho_{tot}(t)=\rho(t)\otimes \Lambda
\end{equation}
where 
\begin{equation}
\rho(t)={\rm tr}_B\{\rho_{tot}(t)\}
\end{equation}
is the usual reduced density matrix of the system. In consequence
\begin{eqnarray}
&&PL_{tot}e^{-iQL_{tot}t}Q\rho_{tot}(0)=PL_{tot}e^{-iQL_{tot}t}Q\rho(0)\otimes \rho_B\\
&&=[S,{\rm tr}_B\{Be^{-iQL_{tot}t}\rho(0)\otimes(\rho_B-\Lambda)\}]\otimes \Lambda\\
&&PL_{tot}P\rho_{tot}(t)=PL_{tot}\rho(t)\otimes \Lambda=[H+S\overline{{\cal B}},\rho(t)]\otimes\Lambda\\
&&QL_{tot}P\rho_{tot}(t)=QL_{tot}\rho(t)\otimes\Lambda=[S(B-\overline{{\cal B}}),\rho(t)\otimes\Lambda],
\end{eqnarray}
where $\overline{{\cal B}}={\rm Tr}_B\{B\Lambda\}$, and it follows that $\rho(t)$ obeys an ensemble of master equations each of the form
\begin{eqnarray}
&&\frac{d\rho(t)}{dt}=-i[H+S\overline{{\cal B}},\rho(t)]-i[S,(M^{(1)}(t)-M^{(2)}(t))\rho(0)]\nonumber \\
&&-\int_0^tdt'~[S,\left(\frac{1}{2}(M^{(3)}(t-t')+M^{(4)}(t-t'))-\overline{{\cal B}}M^{(2)}(t-t')\right)[S,\rho(t')]\nonumber \\
&&+\frac{1}{2}\left(M^{(3)}(t-t')-M^{(4)}(t-t')\right)[S,\rho(t')]_+]\label{ME2}
\end{eqnarray}
where $[S,\rho(t')]_+=S\rho(t')+\rho(t')S$ and
\begin{eqnarray}
M^{(1)}(t)&=&{\rm tr}_B\{Be^{-iQL_{tot}t}\rho_B\}\label{MOP1}\\
M^{(2)}(t)&=&{\rm tr}_B\{Be^{-iQL_{tot}t}\Lambda\}\\
M^{(3)}(t)&=&{\rm tr}_B\{Be^{-iQL_{tot}t}B\Lambda\}\\
M^{(4)}(t)&=&{\rm tr}_B\{Be^{-iQL_{tot}t}\Lambda B\}\label{MOP2}
\end{eqnarray}
are memory operators.

It is doubtful that any choice of $\Lambda$ would lead to a set of $M^{(k)}(t)$ which have simple analytic forms. Numerical evaluation of $M^{(k)}(t)$ would be as difficult as solving the full system plus bath Schr\"{o}dinger equation. Hence Eq. (\ref{ME2}) is exact but essentially unsolveable.

\section{Mean field approximation}

A solvable approximate master equation can be obtained by replacing the system operators $M^{(k)}(t)$
by their scalar mean values
\begin{equation}
\langle M^{(k)}(t)\rangle=\frac{1}{N^2}\sum_{j=1}^{N^2}{\rm Tr}_S\{\chi_j^{\dag}M^{(k)}(t)\chi_j\}
\end{equation}
for $k=1,\dots, 4$ where $\chi_j$ are a complete set of states for the system Liouville Hilbert space and $N$ is the dimension of the usual state space. Obviously this is an uncontrolled approximation that introduces an unknown amount of error. Nor is the approximation unique as we will show in Section 8.

In this mean field approximation Eq. (\ref{ME2}) yields solvable master equations
\begin{eqnarray}
\frac{d\rho(t)}{dt}&=&-i[H+S\overline{{\cal B}},\rho(t)]-iK^{(0)}(t)[S,\rho(0)]\nonumber \\
&-&\int_0^tdt'~[S,K^{(1)}(t-t')[S,\rho(t')]+K^{(2)}(t-t')[S,\rho(t')]_+]\label{ME3}
\end{eqnarray}
where
\begin{eqnarray}
K^{(0)}(t) &=&\langle M^{(1)}(t)\rangle -\langle M^{(2)}(t)\rangle\\
K^{(1)}(t)&=&\frac{1}{2}(\langle M^{(3)}(t)\rangle +\langle M^{(4)}(t)\rangle )-\overline{{\cal B}}~\langle M^{(2)}(t)\rangle\\
K^{(2)}(t)&=&\frac{1}{2}(\langle M^{(3)}(t)\rangle -\langle M^{(4)}(t)\rangle )
\end{eqnarray}
are scalar memory functions that differ depending on the choice of projection operator $\Lambda$. Hence, the error will also be dependent on $\Lambda$.

\subsection{Self-consistent Born approximation}
We will employ an operator $\Lambda$, associated with the  projection operator (\ref{PO}), of the form
\begin{equation}
\Lambda=\rho_B+\sum_{j=1}^{n_p}\eta_j (H_B^j-\overline{H_B^j})\rho_B,\label{PROJ}
\end{equation}
which clearly satisfies conditions (\ref{COND1})-(\ref{COND3}).
The mean values of the operators $M^{(k)}(t)$ can then be approximated\cite{Wilk} using techniques from Random Matrix Theory and a second order self-consistent Born approximation\cite{RMT,RMT2}. The details are discussed in Ref. \cite{Wilk}. One subtlety is that the $W(t)$ memory function of Ref. \cite{Wilk} has the property $W(0)=1$ but this is not true of our $\langle M^{(k)}(t)\rangle$. To facilitate application of techniques from Ref. \cite{Wilk} we define an operation such that each of Eqs. (\ref{MOP1})-(\ref{MOP2}) can be written in the mean as $\langle M^{(k)}(t)\rangle=\widehat{e^{-iQL_{tot}t}}_k$. Application of this identity for the total identity operator ${\cal I}={\cal I}_S\otimes {\cal I}_B$ gives for $k=4$, for example, $\langle {\rm tr}_B\{B {\cal I}_S\otimes {\cal I}_B\Lambda B\}\rangle=\hat{{\cal I}}_4$. Our memory functions can be handled according to the methods of Ref. \cite{Wilk} provided we define $\tilde{W}(z)= \langle \tilde{M}^{(k)}(z)\rangle/\hat{{\cal I}}_k$, as the Laplace tranform of $W(t)$, where each $k$ is treated separately and where $\langle \tilde{M}^{(k)}(z)\rangle$ is the Laplace transform of $\langle M^{(k)}(t)\rangle$.

The application of methods from Ref. \cite{Wilk} then yields
\begin{eqnarray}
\langle M^{(1)}(t)\rangle &=&\bar{B} ~W(\alpha_1,\beta_1,t)\label{M1}\\
\langle M^{(2)}(t)\rangle &=&\bar{{\cal B}} ~W(\alpha_2,\beta_2,t)\\
\langle M^{(3)}(t)\rangle &=&\langle M^{(4)}(t)\rangle=\overline{{\cal B}^2}~W(\alpha_3,\beta_3,t)\label{M3}
\end{eqnarray}
where $\bar{B}={\rm Tr}_B\{B\rho_B\}$, $\overline{{\cal B}^2}={\rm Tr}_B\{B^2\Lambda\}$, and 
\begin{equation}
W(\alpha_k,\beta_k,t)=\frac{1}{\sqrt{\pi}}\sum_{n=0}^{\infty}\frac{\Gamma(\frac{n+1}{2})}{n!}(-\alpha_k t)^n\left(\frac{2}{\beta_k t}\right)^{n/2+1}J_{n/2+1}(\beta_k t)
\end{equation}
where
\begin{eqnarray}
\alpha_k&=&(\widehat{{\cal A}{\cal A}^{\dag}}_k-\widehat{{\cal A}{\cal A}}_k)/\sqrt{\hat{{\cal I}}_k\widehat{{\cal A}{\cal A}^{\dag}}_k}\label{alph}\\
\beta_k&=&(\widehat{{\cal A}{\cal A}^{\dag}}_k+\widehat{{\cal A}{\cal A}}_k)/\sqrt{\hat{{\cal I}}_k\widehat{{\cal A}{\cal A}^{\dag}}_k}.\label{bet}
\end{eqnarray}
[Note that Ref. \cite{Wilk} employs notation like $\langle {\cal A}{\cal A}^{\dag}\rangle$ for an average over system and bath degrees of freedom instead of our notation $\widehat{{\cal A}{\cal A}^{\dag}}_k$ which in general does not denote a full average. Only the system part is a true average in our case.] In Eqs. (\ref{alph}) and (\ref{bet}) ${\cal A}=QL_{tot}$ and the formulas for these parameters $\widehat{{\cal A}{\cal A}}_k$, $\widehat{{\cal A}{\cal A}^{\dag}}_k$ and $\hat{{\cal I}}_k$ for different $k$ are given in Table \ref{table1}. Note again that $\hat{{\cal I}}_k$ is a normalization constant introduced because $\langle M^{(k)}(0)\rangle\neq 1$.

Note also that Eq. (\ref{M3}) implies that 
\begin{eqnarray}
K^{(2)}(t)=0. 
\end{eqnarray}
Consequently the mean field master equation reduces to the simpler form
\begin{eqnarray}
\frac{d\rho(t)}{dt}&=&-i[H+S\overline{{\cal B}},\rho(t)]-iK^{(0)}(t)[S,\rho(0)]\nonumber \\
&-&\int_0^tdt'~K^{(1)}(t-t')[S,[S,\rho(t')]].\label{ME4}
\end{eqnarray}
\begin{center}
\begin{table}
\begin{tabular}{|c|c|c|c|}
\hline
k & $\hat{{\cal I}}_k$ & $\widehat{{\cal A}{\cal A}^{\dag}}_k$ & $\widehat{{\cal A}{\cal A}}_k$ \\
\hline
1 & $\bar{B}$ & $\langle {\rm tr}_B\{B{\cal A}{\cal A}^{\dag}\rho_B\}\rangle$ & $\langle {\rm tr}_B\{B{\cal A}{\cal A}\rho_B\}\rangle$ \\
2 & $\bar{{\cal B}}$ & $\langle {\rm tr}_B\{B{\cal A}{\cal A}^{\dag}\Lambda\}\rangle$ & $\langle {\rm tr}_B\{B{\cal A}{\cal A}\Lambda\}\rangle$ \\
3 &$\overline{{\cal B}^2}$ & $\langle {\rm tr}_B\{B{\cal A}{\cal A}^{\dag}B\Lambda\}\rangle$ & $\langle {\rm tr}_B\{B{\cal A}{\cal A}B\Lambda\}\rangle$ \\
\hline
\end{tabular}
\caption{Parameter formulas for the mean field approximation}
\label{table1}
\end{table}
\end{center}

For $k=1$ one can show that the general definitions of Table 1 yield the explicit forms 
\begin{eqnarray}
&&\hbar^2 \widehat{{\cal A}{\cal A}}_1=\langle [H,[H,\cdot]]\rangle ~~(\bar{B}-\bar{{\cal B}})+\langle [S,[H,\cdot]]\rangle ~~(\overline{B^2}-\overline{{\cal B}^2}-\bar{{\cal B}}(\bar{B}-\bar{{\cal B}}))\nonumber \\
&&+\langle [H,[S,\cdot]]\rangle ~(\overline{B^2}-\bar{B}\bar{{\cal B}})+\langle [S,[S,\cdot]]\rangle ~(\overline{B^3}-\overline{B^2}\bar{{\cal B}}-\overline{B}\overline{{\cal B}^2}+\bar{{\cal B}}^2\bar{B})\label{FORM1}\\
&&\hbar^2 \widehat{{\cal A}{\cal A}^{\dag}}_1=\langle [H,[H,\cdot]]\rangle ~(\bar{B}-\bar{{\cal B}}-\bar{\Lambda}~({\rm Tr}_B\{B\}-{\rm Tr}_B\{{\cal I}_B\}\bar{{\cal B}}))\nonumber \\
&&+(\langle [S,[H,\cdot]]\rangle ~+\langle [H,[S,\cdot]]\rangle ~)(\overline{B^2}-\overline{{\cal B}}\bar{B}-\bar{\Lambda}({\rm Tr}_B\{B^2\}-{\rm Tr}_B\{B\}\bar{{\cal B}}))\nonumber \\
&&+\langle [S,[S,\cdot]]\rangle ~(\overline{B^3}-\overline{B^2}\bar{{\cal B}}-\bar{\Lambda}({\rm Tr}_B\{B^3\}-{\rm Tr}_B\{B^2\}\bar{{\cal B}}))
\end{eqnarray}
where caligraphic averages like $\overline{{\cal B}^3}$ are computed using $\Lambda$, i.e. $\overline{{\cal B}^3}={\rm Tr}_B\{B^3\Lambda\}$, while others like $\overline{B^2}$ are computed using $\rho_B$, i.e. $\overline{B^2}={\rm Tr}_B\{B^2\rho_B\}$.
For $k=2$
\begin{eqnarray}
&&\hbar^2 \widehat{{\cal A}{\cal A}}_2=\langle [H,[S,\cdot]]\rangle ~(\overline{{\cal B}^2}-\bar{{\cal B}}^2)+\langle [S,[S,\cdot]]\rangle ~(\overline{{\cal B}^3}-2\overline{{\cal B}^2}\overline{{\cal B}}+\overline{{\cal B}}^3)\\
&&\hbar^2 \widehat{{\cal A}{\cal A}^{\dag}}_2=\langle [H,[H,\cdot]]\rangle ~ {\rm Tr}_B\{\Lambda^2\} ({\rm Tr}_B\{{\cal I}_B\}\bar{{\cal B}}-{\rm Tr}_B\{B\})\nonumber \\
&&+(\langle [S,[H,\cdot]]\rangle ~+\langle [H,[S,\cdot]]\rangle ~)(\overline{{\cal B}^2}-\overline{{\cal B}}^2-{\rm Tr}_B\{\Lambda^2\}({\rm Tr}_B\{B^2\}-{\rm Tr}_B\{B\}\bar{{\cal B}}))\nonumber \\
&&+\langle [S,[S,\cdot]]\rangle ~(\overline{{\cal B}^3}-\overline{{\cal B}^2}\bar{{\cal B}}-{\rm Tr}_B\{\Lambda^2\}({\rm Tr}_B\{B^3\}-{\rm Tr}_B\{B^2\}\bar{{\cal B}}))
\end{eqnarray}
and for $k=3$
\begin{eqnarray}
&&\hbar^2 \widehat{{\cal A}{\cal A}}_3=\langle [H,[H,\cdot]]\rangle ~(\overline{{\cal B}^2}-\bar{{\cal B}}^2)+\langle [S,[H,\cdot]]\rangle ~(\overline{{\cal B}^3}-2\overline{{\cal B}^2}\bar{{\cal B}}+\bar{{\cal B}}^3)\nonumber \\
&&+\langle [H,[S,\cdot]]\rangle ~(\overline{{\cal B}^3}-\bar{{\cal B}}\overline{{\cal B}^2})+\langle [S,[S,\cdot]]\rangle ~(\overline{{\cal B}^4}-\overline{{\cal B}^3}\bar{{\cal B}}-\overline{{\cal B}^2}^2+\overline{{\cal B}}^2\overline{{\cal B}^2})\nonumber\\
&&+{\rm Tr}_B\{BH_B^2B\Lambda\}-2{\rm Tr}_B\{H_BBH_BB\Lambda\}+{\rm Tr}_B\{H_B^2B^2\Lambda\}\\
&&\hbar^2 \widehat{{\cal A}{\cal A}^{\dag}}_3=\langle [H,[H,\cdot]]\rangle ~(\overline{{\cal B}^2}-\bar{{\cal B}}^2-{\rm Tr}_B\{B\Lambda^2\}({\rm Tr}_B\{B\}-{\rm Tr}_B\{{\cal I}_B\}\bar{{\cal B}}))\nonumber \\
&&+(\langle [S,[H,\cdot]]\rangle ~+\langle [H,[S,\cdot]]\rangle ~)(\overline{{\cal B}^3}-\overline{{\cal B}^2}\bar{{\cal B}}-{\rm Tr}_B\{B\Lambda^2\}({\rm Tr}_B\{B^2\}-{\rm Tr}_B\{B\}\bar{{\cal B}}))\nonumber \\
&&+\langle [S,[S,\cdot]]\rangle ~(\overline{{\cal B}^4}-\overline{{\cal B}^3}\bar{{\cal B}}-{\rm Tr}_B\{B\Lambda^2\}({\rm Tr}_B\{B^3\}-{\rm Tr}_B\{B^2\}\bar{{\cal B}}))\nonumber \\
&&+{\rm Tr}_B\{BH_B^2B\Lambda\}-2{\rm Tr}_B\{H_BBH_BB\Lambda\}+{\rm Tr}_B\{H_B^2B^2\Lambda\}.\label{FORM6}
\end{eqnarray}
Note additionally that with projection operator (\ref{PROJ})
\begin{eqnarray}
&&\overline{{\cal B}^l}=\overline{B^l}+\sum_{j=1}^{n_p}\eta_j(\overline{B^lH_B^j}-\overline{B^l}\overline{H_B^j})\\
&&\bar{\Lambda}=\overline{\rho_B}+\sum_{j=1}^{n_p}\eta_j(\overline{\rho_BH_B^j}-\overline{\rho_B}\overline{H_B^j})
\end{eqnarray}
and other quantities $\bar{X}$ can be similarly computed via $\bar{X}={\rm Tr}\{X\rho_B\}$ (e.g. $\overline{\rho_B}={\rm Tr}_B\{\rho_B^2\}$).

All parameters for the mean field master equation are now well defined for a specific choice of $\Lambda$. We could in fact compare the results from master equation (\ref{ME4}) to exact results computed using the methods of Reference \cite{TW}.
The bath traces and averages in the above formulas, as in Ref. \cite{TW}, would be computed in the eigenbasis of $H_B$ and would include only the first $n_B$ eigenstates. Thus, for example, ${\rm Tr}_B\{{\cal I}_B\}=n_B$. The parameter $n_B$ is determined by the temperature and it is assumed that states higher in energy are essentially unpopulated.

However, master equation (\ref{ME4}) may not preserve positivity and may not correctly equilibrate. Hence, we will instead try to match the time-scales of (\ref{ME4}) with a more abstract theory which does preserve positivity and does equilibrate. We will primarily focus on matching the memory functions $K^{(1)}(t)$ for the two theories. The abstract theory is discussed next.

\section{Positivity requirements and equilibration}

We consider a master equation with the same general structure as Eq. (\ref{ME4}) but with potentially different memory functions. Our first goal, in Section 4.1, will be to show that memory functions exist for which the density matrix has a stochastic decomposition and therefore preserves positivity. By stochastic decomposition we mean that $\rho(t)$ can be written as an average
\begin{equation}
\rho(t)=M[|\psi(t)\rangle\langle\psi(t)|]\label{SDECOMP}
\end{equation}
where $M[\cdot]$ is a mean evaluated over different stochastically evolving wavefunctions $|\psi(t)\rangle $.

Below in Section 4.2 we will examine the circumstances under which equilibration is possible.

\subsection{Stochastic decomposition}

Here we generalize the proof from Ref. \cite{WW3} to the case of inhomogeneous master equations. We also relax one unnecessary restriction imposed in Ref. \cite{WW3}.

Following Ref. \cite{WW3} we assume that we may write the Laplace transform $\tilde{K}^{(1)}(z)$ of the memory function $K^{(1)}(t)$ in the form
\begin{eqnarray}
\frac{1}{\tilde{K}^{(1)}(z)}=\frac{1}{K^{(1)}(0)}~[\tilde{V}(z)+z-\lambda]\label{VEQ}
\end{eqnarray}
and we will require that $V(t)$, the inverse Laplace transform of $\tilde{V}(z)$, obey the inequalities
\begin{eqnarray}
&&V(t)\geq 0\label{VEQ2}\\
&&V'(t)=dV(t)/dt\leq 0.\label{VEQ3} 
\end{eqnarray}
We will also assume that the effects of the drift governed by $L=[H+\bar{{\cal B}}S,\cdot]$ and the dissipation governed by ${\cal L}=[S,[S,\cdot]]$ can be treated separately with the overall dynamics obtained via a Trotter product formula\cite{W1}. Thus, we will focus on the evolution equation 
\begin{eqnarray}
\frac{d\rho(t)}{dt}&&=-iK^{(0)}(t)[S,\rho(0)]\nonumber \\
&&-\int_0^tdt'~K^{(1)}(t-t') [S^2\rho(t')+\rho(t')S^2-2S\rho(t')S]\label{EVO}
\end{eqnarray}
which after a Laplace transformation can be written as 
\begin{eqnarray}
\frac{1}{\tilde{K}^{(1)}(z)}~(z\tilde{\rho}(z)-\rho(0))&&=-i\left(\tilde{R}(z)+\frac{K^{(0)}(0)}{K^{(1)}(0)}\right)[S,\rho(0)]\nonumber\\
&&-[S^2\tilde{\rho}(z)+\tilde{\rho}(z)S^2-2S\tilde{\rho}(z)S]
\end{eqnarray}
where we have defined the Laplace transform $\tilde{R}(z)$ of a function $R(t)$ via
\begin{equation}
\tilde{R}(z)+\frac{K^{(0)}(0)}{K^{(1)}(0)}=\frac{\tilde{K}^{(0)}(z)}{\tilde{K}^{(1)}(z)}.\label{RDEF}
\end{equation}
Using Eq. (\ref{VEQ}) and the fact that (\ref{EVO}) implies
\begin{equation}
\frac{d\rho(t)}{dt}\mid_{t=0}=-iK^{(0)}(0)[S,\rho(0)],\label{init2}
\end{equation}
we then have
\begin{eqnarray}
&&z(z\tilde{\rho}(z)-\rho(0))-\frac{d}{dt}\rho(t)\mid_{t=0}-\lambda (z\tilde{\rho}(z)-\rho(0))\nonumber \\
&&=-iK^{(1)}(0)\tilde{R}(z)[S,\rho(0)]\nonumber\\
&&-\tilde{V}(z)(z\tilde{\rho}(z)-\rho(0))-K^{(1)}(0)[S^2\tilde{\rho}(z)+\tilde{\rho}(z)S^2-2S\tilde{\rho}(z)S].\label{LEVO}
\end{eqnarray}
It follows that we may write
\begin{eqnarray}
&&\frac{d^2\rho(t)}{dt^2}-\lambda \frac{d\rho(t)}{dt}=-\int_0^tdt'~V(t-t')\frac{d\rho(t')}{dt'}\nonumber \\
&&-iK^{(1)}(0)R(t)[S,\rho(0)]-K^{(1)}(0)[S^2\rho(t)+\rho(t)S^2-2S\rho(t)S]\\
&&=-V(0)\rho(t)+V(t)\rho(0)-\int_0^tdt' ~V'(t-t')\rho(t')\nonumber \\
&&-iK^{(1)}(0)R(t)[S,\rho(0)]-K^{(1)}(0)[S^2\rho(t)+\rho(t)S^2-2S\rho(t)S],\label{EVO2}
\end{eqnarray}
by inverse Laplace transformation of Eq. (\ref{LEVO}). Noting that
\begin{equation}
e^{\lambda t/2}\frac{d^2[\rho(t)e^{-\lambda t/2}]}{dt^2}=\frac{d^2\rho(t)}{dt^2}-\lambda \frac{d\rho(t)}{dt}+(\lambda/2)^2\rho(t)
\end{equation}
and substituting into Eq. (\ref{EVO2}) we finally have 
\begin{eqnarray}
&&\frac{d^2[\rho(t)e^{-\lambda t/2}]}{dt^2}=[\lambda^2/4-V(0)]\rho(t)e^{-\lambda t/2}+V(t)e^{-\lambda t/2}\rho(0)\nonumber\\
&&-\int_0^tdt' V'(t-t')e^{-\lambda (t-t')/2}\rho(t')e^{-\lambda t'/2}-iK^{(1)}(0)R(t)e^{-\lambda t/2}[S,\rho(0)]\nonumber \\
&&-K^{(1)}(0)[S^2\rho(t)e^{-\lambda t/2}+\rho(t)e^{-\lambda t/2}S^2-2S\rho(t)e^{-\lambda t/2}S]\label{RHOE}
\end{eqnarray}
which has a stochastic decomposition provided that Eqs. (\ref{VEQ})-(\ref{VEQ3}), $\lambda^2/4\geq V(0)$,  as well as other constraints discussed below are satisfied. Note that Ref. \cite{WW3} required that $\lambda^2/4=V(0)$ which we will see is unnecessary.

To avoid dealing with specific initial conditions $\rho(0)$ we employ the Hadamard representation\cite{WW1}.  The Hadamard representation of the propagator is defined via $\langle s| \rho(t)|s'\rangle=U_{s,s'}(t)\langle s| \rho(0)|s'\rangle$ where $S|s\rangle=s|s\rangle$. If the eigenvalues of $U_{s,s'}(t)$ are positive then $\rho(t)$ will be positive semidefinite for all positive semidefinite initial densities\cite{WW1}. The propagator can also be calculated from $U_{s,s'}(t)=M[u_s(t)u_{s'}^*(t)]$ using Eq. (\ref{SDECOMP}) where $\langle s|\psi(t)\rangle=u_s(t)\langle s|\psi(0)\rangle$ and $u_s(t)$ is again defined in the Hadamard sense. Rather than deal directly with $u_s(t)$ and $U_{s,s'}(t)$ we defined instead $v_s(t)=u_s(t)e^{-\lambda t/4}$ and ${\cal V}_{s,s'}(t)=U_{s,s'}(t)e^{-\lambda t/2}$ where ${\cal V}_{s,s'}(t)=M[v_s(t)v_{s'}^*(t)]$. 

Let $\sigma_t=-R(t)/|R(t)|$ then provided that $|R(t)|e^{-\lambda t/2}$ is monotonically decreasing, the It\^{o} equations\cite{BP}
\begin{eqnarray}
&&dv_s(t)=i\sigma_t\sqrt{K^{(1)}(0)}~sv_s(t)~dt+y_s(t)~dw_t^{(1)}\nonumber \\
&&+\sqrt{|R(t)|e^{-\lambda t/2}\sqrt{K^{(1)}(0)}}dw_t^{(2)} \\
dy_s(t)&=&-i\sigma_t\sqrt{K^{(1)}(0)}~sy_s(t)~dt+\sqrt{V(t)}e^{-\lambda t/4}~dw_t^{(3)}\nonumber \\
&&+\int_0^tdw_{t'}^{(4)}\sqrt{-V'(t-t')}e^{-\lambda(t-t')/4}v_s(t') ~dw_t^{(5)}\nonumber \\
&&+\sqrt{-\frac{d}{dt}\left(|R(t)|e^{-\lambda t/2}\sqrt{K^{(1)}(0)}\right)}dw_t^{(6)}\nonumber \\
&&+\sqrt{\lambda^2/4-V(0)}v_s(t) ~dw_t^{(7)}
\label{SD}
\end{eqnarray}
yield the equations
\begin{eqnarray}
\frac{d}{dt}\overline{v_s(t)v_{s'}^*(t)}&=&i\sigma_t\sqrt{K^{(1)}(0)}(s-s')\overline{v_s(t)v_{s'}^*(t)}\nonumber \\
&+&\overline{y_s(t)y_{s'}^*(t)}+|R(t)|e^{-\lambda t/2}\sqrt{K^{(1)}(0)}\label{MEV1}\\
\frac{d}{dt}\overline{y_s(t)y_{s'}^*(t)}&=&-i\sigma_t\sqrt{K^{(1)}(0)}(s-s')\overline{y_s(t)y_{s'}^*(t)}+ V(t)e^{-\lambda t/2}\nonumber \\
&+&\int_0^tdt'[-V'(t-t')]e^{-\lambda(t-t')/2}\overline{v_s(t')v_{s'}^*(t')}\nonumber \\
&+&(\lambda^2/4-V(0))\overline{v_s(t)v_{s'}^*(t)}\nonumber \\
&-&\frac{d}{dt}\left(|R(t)|e^{-\lambda t/2}\sqrt{K^{(1)}(0)}\right)\label{MEV2}
\end{eqnarray}
in the mean. Note that $dw_t^{(j)}$ are real independent Wiener processes\cite{BP}. Provided that the sign of $R(t)$ does not change,
taking the derivative of (\ref{MEV1}) and using (\ref{MEV2}) then gives the equation 
\begin{eqnarray}
\frac{d^2{\cal V}_{s,s'}(t)}{dt^2}&&=(\lambda^2/4-V(0)){\cal V}_{s,s'}(t)-iR(t)K^{(1)}(0)e^{-\lambda t/2}(s-s')\nonumber \\
&&+ V(t)e^{-\lambda t/2}-\int_0^tdt' V'(t-t')e^{-\lambda (t-t')/2}{\cal V}_{s,s'}(t')\nonumber \\
&&-K^{(1)}(0)(s-s')^2{\cal V}_{s,s'}(t)\label{SD2}
\end{eqnarray}
which is the Hadamard representation of Eq. (\ref{RHOE}). Initial conditions $v_s(0)=1$ guarantee that ${\cal V}_{s,s'}(0)=1$. Hence, we find that
\begin{eqnarray} 
&&\frac{d}{dt}\overline{v_s(t)v_{s'}^*(t)}|_{t=0}=i\sigma_0\sqrt{K^{(1)}(0)}(s-s')+y_s(0)y_{s'}^*(0)\nonumber \\
&&+|R(0)|\sqrt{K^{(1)}(0)}
\end{eqnarray}
which we require to equal 
\begin{eqnarray} 
\frac{d}{dt}\overline{v_s(t)v_{s'}^*(t)}|_{t=0}=-iK^{(0)}(0)(s-s') -\lambda/2
\end{eqnarray}
in order to match the original initial condition (\ref{init2}). Thus, it appears that we must choose
\begin{eqnarray}
&&K^{(0)}(0)=-\sigma_0\sqrt{K^{(1)}(0)}\label{VEQ4}\\
&&-\lambda/2 \geq |R(0)|\sqrt{K^{(1)}(0)}\label{VEQ5}
\end{eqnarray}
so that we can pick initial values $y_s(0)=\sqrt{-\lambda/2 -|R(0)|\sqrt{K^{(1)}(0)}}$. This clearly implies that
\begin{eqnarray}
\lambda < 0.
\end{eqnarray}

Since the Hadamard propagator has a stochastic decomposition it is positive semidefinite and so $\rho(t)$ will be positive semidefinite. This completes the proof of positivity for $\rho(t)$ provided that a memory function satisfying (\ref{VEQ})-(\ref{VEQ3}) and (\ref{VEQ4})-(\ref{VEQ5}), $\lambda^2/4\geq V(0)$, and $\sigma_t$ constant and $|R(t)|e^{-\lambda t/2}$ monotonically decreasing, can be found. We will discuss an appropriate set of memory functions in Section 5.

\subsection{Equilibration}

The following argument generalizes results from Ref. \cite{WW3} to the case of an inhomogeneous master equation.

Consider a non-Markovian Kossakowsi-Lindblad\cite{dsg} (NMKL) master equation of integrodifferential form\cite{Bud,WW3,WW1,W1} which employs a scalar memory function $K(t)$ and has a single Kossakowski-Lindblad dissipation operator constructed from the system interaction operator $S$:
\begin{eqnarray}
\frac{d\rho(t)}{dt}&&=F(t)-i[H+\overline{{\cal B}}S,\rho(t)]\nonumber \\
&&-\int_0^tdt'~K(t-t') [S^2\rho(t')+\rho(t')S^2-2S\rho(t')S].\label{NMME}
\end{eqnarray}
Here $F(t)$ represents the inhomogeneous term.

For additional notational simplicity we will introduce 
operators $L=[H+\overline{{\cal B}}S,\cdot]$ and ${\cal L}=[S,[S,\cdot]]$. Laplace transforming
equation (\ref{NMME}) yields
\begin{eqnarray}
z\tilde{\rho}(z)-\rho(0)=\tilde{F}(z)-iL\tilde{\rho}(z)-\tilde{K}(z) {\cal L}\tilde{\rho}(z)
\end{eqnarray}
from which one can then show that
\begin{equation}
\tilde{\rho}(z)=(z+iL+\tilde{K}(z) {\cal L})^{-1}[\rho(0)+\tilde{F}(z)]\label{RHOT}
\end{equation}
where $\tilde{\rho}(z)=\int_0^{\infty}dt~e^{-zt}\rho(t)$ is the Laplace transformed reduced density matrix, $\tilde{K}(z)$ is the transform of the memory function and $\tilde{F}(z)$ is the transform of the inhomogeneous term. We then define an operator
\begin{eqnarray}
G(z)=(z+iL+\tilde{K}(z) {\cal L})^{-1}
\end{eqnarray}
so that $\tilde{\rho}(z)=G(z)[\rho(0)+\tilde{F}(z)]$.

Defining an operator $G_0(z)=(z+iL)^{-1}$, and letting $|n\rangle$ and $E_n$ denote the complete set of eigenvectors and eigenvalues of $H+\overline{{\cal B}}S$, it then follows that 
\begin{eqnarray}
\lim_{z\rightarrow 0}zG_0(z)|n\rangle\langle n|&=&|n\rangle\langle n| \\
\lim_{z\rightarrow 0}zG_0(z)|n\rangle\langle m|&=&0.
\end{eqnarray}
[We have assumed that the spectrum of $H+\overline{{\cal B}}S$ is discrete. Generalization to
the case where all or part is continuous is straightforward.]
In fact $\lim_{z\rightarrow 0}zG_0(z)=\Pi_0$ where $\Pi_0=\sum_n|n\rangle\langle n|{\rm Tr}_S\{|n\rangle\langle n|\cdot\}$ is a projection operator. 

It then follows that 
\begin{eqnarray}
G(z)&=&(G_0(z)^{-1}+ \tilde{K}(z)  {\cal L})^{-1}\\
&=&[G_0(z)^{-1}(1+G_0(z)\tilde{K}(z)  {\cal L})]^{-1}\\
&=&(1+G_0(z)\tilde{K}(z)  {\cal L})^{-1}G_0(z)
\end{eqnarray}
Hence from Eq. (\ref{RHOT}) we obtain
\begin{eqnarray}
\rho(\infty)&=&\lim_{z\rightarrow 0}zG(z)[\rho(0)+\tilde{F}(z)]\\
&=&\lim_{z\rightarrow 0}(1+G_0(z) \tilde{K}(z)  {\cal L})^{-1}zG_0(z)[\rho(0)+\tilde{F}(z)]\\
&=&\lim_{z\rightarrow 0}(1+G_0(z) \tilde{K}(z)  {\cal L})^{-1} \Pi_0[\rho(0)+\tilde{F}(z)]\\
&=&\lim_{z\rightarrow 0}\left(1+zG_0(z) (\tilde{K}(z)/z)  {\cal L}\right)^{-1} \Pi_0[\rho(0)+\tilde{F}(z)].
\end{eqnarray}
Thus, if the limit 
\begin{equation}
\lim_{z\rightarrow 0}\frac{\tilde{K}(z)}{z}=\kappa\label{C1}
\end{equation}
exists and 
\begin{equation}
\lim_{z\rightarrow 0}\tilde{F}(z)=0\label{CONDF}
\end{equation}
then
\begin{eqnarray}
\rho(\infty)=(1+\kappa \Pi_0 {\cal L})^{-1} \Pi_0\rho(0)\label{Asy}
\end{eqnarray}
and $\rho(\infty)$ satisfies $\Pi_0\rho(\infty)=\rho(\infty)$. Perhaps this is more clearly seen by trivially rewriting (\ref{Asy}) as $\rho(\infty)=\Pi_0\rho(0)-\kappa \Pi_0 {\cal L}(1+\kappa \Pi_0 {\cal L})^{-1} \Pi_0\rho(0)$ and recalling that $\Pi_0^2=\Pi_0$. In addition, since $\Pi_0\rho(0)=\sum_{n}\langle n|\rho(0)|n\rangle ~|n\rangle\langle n|$ the long time limit depends at most on the diagonal elements of the initial density matrix. Condition (\ref{CONDF}) is necessary to prevent dependence on the off-diagonal elements of the initial density matrix. Thus partial equilibration is possible in this inhomogeneous NMKL model. Exact calculations for some model systems equilibrate in a similar manner\cite{T1,T2}.

\section{SME memory functions}

To find appropriate memory functions $K^{(0)}(t)$ and $K^{(1)}(t)$ we start with a fairly general polynomial model for $\tilde{K}^{(1)}(z)$ which obeys the basic constraints discussed above and in Ref. \cite{WW3}. Specifically, we assume that
\begin{equation}
\tilde{K}^{(1)}(z)=K^{(1)}(0)\frac{z(z+\beta)}{z^3+\mu z^2+\nu z +\gamma}.\label{MEMZ}
\end{equation}
It is necessary that the polynomial in the denominator be of one degree higher than that in the numerator so that a non-zero initial value $K^{(1)}(0)$ is guaranteed for the memory function. The factor of $z$ in the numerator guarantees a non-zero limit $\kappa$ in Eq. (\ref{C1}) provided that $\beta$ and $\gamma$ are non-zero. This also guarantees that the memory function is integrable and so has a proper Markovian limit. For non-zero $\beta$ either $\nu$ or $\gamma$ must be non-zero to guarantee that memory is of finite duration.

Given the memory function (\ref{MEMZ}) it then follows that we must have
\begin{eqnarray}
\lambda&=&\beta-\mu\\
\tilde{V}(z)&=&[\nu-\mu\beta +\beta^2+\frac{\gamma}{z}]\frac{1}{z+\beta}.\label{Vmod}
\end{eqnarray}
The $\kappa$ of Eq. (\ref{C1}) can also be obtained as $\kappa=K^{(1)}(0)\beta/\gamma$.
Now $V(t)$ will be positive if $\tilde{V}(z)$ is completely monotone\cite{Fell}. If $\nu-\mu\beta +\beta^2> 0$ and $\gamma> 0$ then the first factor in Eq. (\ref{Vmod}) will be completely monotone. If $\beta > 0$ then the second factor will also be completely monotone. The product of completely monotone functions is completely monotone\cite{Fell}. Hence, $\tilde{V}(z)$ is completely monotone and constraint (\ref{VEQ2}) is satisfied, i.e. $V(t)\geq 0$. 

Note also that 
\begin{equation}
V(0)=\lim_{z\rightarrow \infty}z\tilde{V}(z)=\nu-\mu\beta +\beta^2.
\end{equation}
Denote $\Delta (t)=-V'(t)$, which must be positive by (\ref{VEQ3}), and its Laplace transform will be $\tilde{\Delta}(z)=V(0)-z\tilde{V}(z)$. Inserting Eq. (\ref{Vmod}) gives
\begin{equation}
\tilde{\Delta}(z)=\frac{\beta(\nu-\mu\beta +\beta^2)}{z+\beta}
\end{equation} 
which is clearly completely monotone and so $\Delta(t)$ will be positive.

For $K^{(0)}(t)$ we know that (\ref{VEQ4})-(\ref{VEQ5}) must be obeyed (see section 4.1), $R(t)e^{-\lambda t/2}$ defined via (\ref{RDEF}) must not change sign and should be decreasing in magnitude (see section 4.1), and additionally (\ref{CONDF}) must be satisfied. The model 
\begin{equation}
\tilde{K}^{(0)}(z)=K^{(0)}(0)\frac{z(z+\alpha)}{z^3+\mu z^2+\nu z +\gamma}\label{MEMZ2}
\end{equation}
satisfies all of these constraints. Clearly it follows from definition (\ref{RDEF}), as well as (\ref{MEMZ}) and (\ref{MEMZ2}), that
\begin{equation}
\tilde{R}(z)=\frac{K^{(0)}(0)}{K^{(1)}(0)}\left(\frac{z+\alpha}{z+\beta}-1\right)
\end{equation}
so that
\begin{equation}
R(t)=\frac{(\alpha-\beta)K^{(0)}(0)}{K^{(1)}(0)}\e^{-\beta t}
\end{equation}
and 
\begin{equation}
R(t)e^{-\lambda t/2}=\frac{(\alpha-\beta)K^{(0)}(0)}{K^{(1)}(0)}\e^{-(3\beta-\mu) t/2}
\end{equation}
which is monotonically decreasing in magnitude provided that
\begin{eqnarray}
3\beta > \mu,\label{BMCOND}
\end{eqnarray}
and it does not change sign.

Note also that
\begin{eqnarray}
\lim_{z\rightarrow 0}\tilde{K}^{(0)}(z)&&=\lim_{z\rightarrow 0}K^{(0)}(0)\frac{z(z+\alpha)}{z^3+\mu z^2+\nu z +\gamma}\\
&&=0,
\end{eqnarray}
and since $\tilde{F}(z)=-i[S,\rho(0)]\tilde{K}^{(0)}(z)$ it follows that (\ref{CONDF}) is satisfied.

Finally, conditions (\ref{VEQ4})-(\ref{VEQ5}) require that
\begin{eqnarray}
&&|K^{(0)}(0)|=\sqrt{K^{(1)}(0)}\\
&&\beta \geq \alpha \geq \beta+\lambda/2.
\end{eqnarray}
We arbitrarily set $\alpha=\beta+\lambda/4$ and $K^{(0)}(0)=\sqrt{K^{(1)}(0)}$. In the calulations reported below $K^{(0)}(t)$ is insensitive to the value chosen for $\alpha$, and the reduced density matrix elements are insensitive to $K^{(0)}(t)$.

\section{Nitrogen-Vacancy model}

The native Hamiltonian is that of an $S=1$ electronic spin associated with a single {\rm NV} impurity center in diamond in its ground electronic state, and
it takes the form\cite{NKPPK}
\begin{eqnarray}
H=h_xS_X+DS_X^2+E(S_X^2-S_Y^2)
\end{eqnarray}
where the magnetic field is along the $x$-direction, $h_x=g_e\mu_eB_x$, and $S_X$, $S_Y$ and $S_Z$ are operator components of the electron spin
with the form\cite{NKPPK}
\[S_X=\frac{1}{\sqrt{2}}\left(\begin{array}{ccc}
0 & 1 & 0 \\
1 & 0 & 1 \\
0 & 1 & 0 \end{array} \right) \]
\[S_Y=\frac{i}{\sqrt{2}}\left(\begin{array}{ccc}
0 & -1 & 0 \\
1 & 0 & -1 \\
0 & 1 & 0 \end{array} \right) \]
\[S_Z=\left(\begin{array}{ccc}
1 & 0 & 0 \\
0 & 0 & 0 \\
0 & 0 & -1 \end{array} \right). \]
The bath consists of 18 $^{13}$C impurity nuclear spins with Hamiltonian\cite{Dob}
\begin{eqnarray}
H_B=h_x^{(0)}\sum_j I_x^{(j)}+(\beta/C)\sum_{j=1}^{17}\sum_{k=j+1}^{18}C_{j,k}( 3I_z^{(j)}I_z^{(k)}-{\bf I}^{(j)}\cdot{\bf I}^{(k)})
\end{eqnarray}
and\cite{Dob2} $h_x^{(0)}=g_n\mu_nB_x$,
\begin{eqnarray}
C_{j,k}=[1-3(z_j-z_k)^2/|{\bf r}_j-{\bf r}_k|^2]/|{\bf r}_j-{\bf r}_k|^3,
\end{eqnarray}
where $C^2=\sum_{j=1}^{17}\sum_{k=j+1}^{18} C_{j,k}^2$. Here $I^{(j)}_x$, $I^{(j)}_y$ and $I^{(j)}_z$, denote the components of the vector spin ${\bf I}^{(j)}$ for the $j$th nucleus, and have the form of their corresponding Pauli matrices multiplied by $I=1/2$. The factor of $1/C$ is required to correct the units and guarantee a proper thermodynamic limit. The impurity locations were selected randomly from a spherical integer lattice with a radius of 5 units. This means about 4 \% of the lattice sites are occupied by $^{13}$C impurities whereas in real diamond the natural abundance is 1.1 \% \cite{NKPPK}. Of course we should really choose the correct diamond lattice locations\cite{NKPPK} but our model has more serious flaws. 

The system-bath coupling should consist of a complete tensor coupling all system spin components with all
bath spin components\cite{NKPPK}, but the present formulation of the master equation can only handle a single
system coupling operator. We thus choose arbitrarily to include only the $x$-components, i.e., we pick a system-bath coupling operator of the form
\begin{equation}
H_{SB}=S_X\otimes (1/A)\sum_k A_k (A_{XX}I_x^{(k)}+A_{XY}I_y^{(k)}+A_{XZ}I_z^{(k)})
\end{equation}
where\cite{Dob2}
\begin{equation}
A_k=(1-3 z_k^2/|{\bf r}_k|^2)/|{\bf r}_k|^3
\end{equation}
and $A^2=\sum_{k=1}^{18} A_k^2$.
Hence, our model is not completely correct at present. Until we generalize the abstract theory to multiple system coupling operators this is the best we can do. The total Hamiltonian is thus of the form (\ref{HAM1}) with $S=S_X$ and $B=(1/A)\sum_k A_k (A_{XX}I_x^{(k)}+A_{XY}I_y^{(k)}+A_{XZ}I_z^{(k)})$ with parameter values given in Table 2 for a magnetic field of 11 G\cite{NKPPK,Dob}.

\begin{center}
\begin{table}
\begin{tabular}{|c|c|c|}
\hline
NV model Parameter & Value (in GHz)& Reference\\
\hline
$h_x$ & .194 & \cite{NKPPK}\\
$D$ & 2.88 &\cite{NKPPK}\\
$E$ & .1 &\cite{Note}\\
$A_{XX}$ & .2 & \cite{NKPPK}\\
$A_{XY}$ & .02 &\cite{NKPPK}\\
$A_{XZ}$ & .02& \cite{NKPPK}\\
$h_x^{(0)}$ & $1.08\times 10^{-3}$ & \cite{Guru}\\
$\beta$ & $4.52\times 10^{-5}$ & \cite{Guru}\\
$k_BT$ & .0003 & \cite{Guru}\\
\hline
\end{tabular}
\caption{Numerical values of NV model parameters for $B_x=$ 11 G}
\label{table2}
\end{table}
\end{center}

\begin{center}
\begin{table}
\begin{tabular}{|c|c|}
\hline
Mean field memory function parameter & Value \\
\hline
$\overline{{ B}}$ & $9.3513\times 10^{-2}$\\
$\overline{{\cal B}}$ & $9.3276\times 10^{-2}$\\
$\overline{{\cal B}^2}$ & $9.8692\times 10^{-3}$\\
$\alpha_1$ & 1.4111\\
$\beta_1$ & 1.4259\\
$\alpha_2$ & 1.3935\\
$\beta_2$ & 1.3951\\
$\alpha_3$ & 1.1953\\
$\beta_3$ & 1.7843\\
\hline
\end{tabular}
\caption{Numerical values of parameters for memory functions of mean field master equation}
\label{table3}
\end{table}
\end{center}

\begin{center}
\begin{table}
\begin{tabular}{|c|c|}
\hline
SME parameter & Value \\
\hline
$\beta$ & 45.9675\\
$\mu$ & 46.4375\\
$\nu$ & 21.6505\\
$\gamma$ & 106.1616\\
$\overline{{\cal B}}$ & $9.3276\times 10^{-2}$\\
$K^{(1)}(0)$ & $1.1665\times 10^{-3}$\\
\hline
\end{tabular}
\caption{Numerical values of parameters for memory functions of SME}
\label{table4}
\end{table}
\end{center}

The initial density matrix was chosen to take the form (\ref{INIT}) with
\begin{eqnarray}
|\psi(0)\rangle = \frac{1}{\sqrt{5}}[\sqrt{3}|-\rangle +i|0\rangle +|+\rangle]
\end{eqnarray}
where $|-\rangle$, $|0\rangle$ and $|+\rangle$ denote eigenstates of $S_X$ with corresponding eigenvalues $-1$, 0 and 1.

\subsection{Determination of parameters of SME}

Equations (\ref{FORM1})-(\ref{FORM6}) involve quantities
\begin{eqnarray}
&&\langle [H,[H,\cdot]]\rangle=(D+E)^2/3+4h_x^2/3+(D-3E)^2/9\\
&&\langle [H,[S,\cdot]]\rangle=4h_x/3\\
&&\langle [S,[H,\cdot]]\rangle=4h_x/3\\
&&\langle [S,[S,\cdot]]\rangle=4/3
\end{eqnarray}
which can be evaluated given $S$ and $H$ for the NV model. All other parameters including the $\eta_j$ of (\ref{PROJ}) for $j=1,\cdots,n_p$ and the $\beta$, $\mu$, $\nu$ and $\gamma$ of (\ref{MEMZ}) were set during a simulated annealing\cite{SIMANN} process where the parameters were optimized so that the mean field memory function $K^{(1)}(t)$ of Section 3 and the formal mathematical model of Section 5 match as well as is possible. We picked $n_p=10$ and required $|\eta_j|\leq 300$. For the parameters of $K^{(1)}(t)$ we defined variables $X_1\cdots,X_4$ via
\begin{eqnarray}
\beta&=&X_1\\
\mu&=&\beta+X_2+2\sqrt{X_3}\\
\nu&=&X_3+\mu\beta-\beta^2\\
\gamma&=& X_4
\end{eqnarray}
and we required $0\leq X_k\leq 200$. The function
\begin{equation}
f(\eta_1,\dots,\eta_{10},X_1,\dots,X_4)=\int_0^{30}dt'~[K^{(1)}_{MF}(t')-K^{(1)}_{SME}(t')]^2,
\end{equation}
where $K^{(1)}_{MF}(t)$ is the mean field memory function calculated via (\ref{M1})-(\ref{M3}), and $K^{(1)}_{SME}(t)$ is the SME memory function based the polynomial model of Section 5, was minimized using the program {\it simann.f} of Ref. \cite{SIMANN}.

The optimal memory function parameter values are given in Table 3 for the mean field master equation and Table 4 for the SME. Note that conditions $3\beta >\mu$, $\beta > 0$, $\gamma > 0$, $V(0)=\nu-\mu\beta+\beta^2=4.5775\times 10^{-2} > 0$, $\lambda=-0.4700<0$, and $\lambda^2/4-V(0)=9.45\times 10^{-3}\geq 0$ are all satisfied.

The resulting mean field $K^{(1)}(t)$ (solid curve) and the polynomial model (dashed curve) are compared in Figure 1. The agreement is quite good. Note that $K^{(1)}(t)$ decays to zero after about 20 time units or about $2\times 10^{-8}$ s. We will however be exploring dynamics for times as long as .1 ms. One might be tempted to conclude that a Markovian formulation would work for this problem. However, we know\cite{THERM1,THERM2} that the Markovian form cannot capture equilibration correctly in general unless the coupling operator takes a very restrictive form. 

Our model for $K^{(0)}(t)$ (dashed curve) shown in Figure 2 is roughly similar to its mean field counterpart (solid curve) except at very short time. A higher order polynomial model would probably improve the agreement. 

\subsection{Numerics}

The exact equations were obtained using the methods of Ref. \cite{TW}. A temperature of $k_BT=.0003$ GHz, or $T=1.44\times 10^{-5}$ K, was chosen for which a total of $n_B=20$ bath eigenstates make significant contributions. (Temperatures an order of magnitude lower can be achieved in practice\cite{Guru}.) All exact calculations and SME parameter calculations were carried out in this eigenbasis. 

The SME (\ref{ME4}) can be mapped to a set of four differential equations by finding the roots of the polynomial $z^3+\mu z^2+\nu z+\gamma=0$. From the three roots $z_k$ one can find $K^{(1)}(t)=\sum_{k=1}^3a_k e^{z_kt}$. Then defining $\Omega_k(t)=\int_0^tdt'~a_k e^{z_k(t-t')}[S,[S,\rho(t')]]$ we obtain
\begin{eqnarray}
&&\frac{d\rho(t)}{dt}=-i[H+S\overline{{\cal B}},\rho(t)]-iK^{(0)}(t)[S,\rho(0)]-\sum_{k=1}^3\Omega_k(t)\\
&&\frac{d\Omega_k(t)}{dt}=a_k[S,[S,\rho(t)]]+z_k\Omega_k(t)
\end{eqnarray}
for $k=1,\dots,3$ with initial conditions $\Omega_k(0)=0$.
All ordinary differential equations were solved using standard Runge-Kutta algorithms\cite{DOP853,HW}.

\begin{figure}
\centering
\includegraphics[scale=0.75]{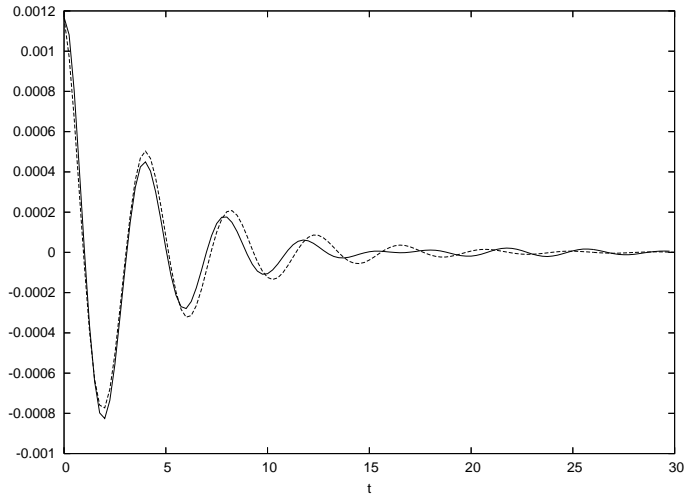}
\caption{$K^{(1)}(t)$ vs. $t$}
\label{fig1}
\end{figure}

\begin{figure}
\centering
\includegraphics[scale=0.75]{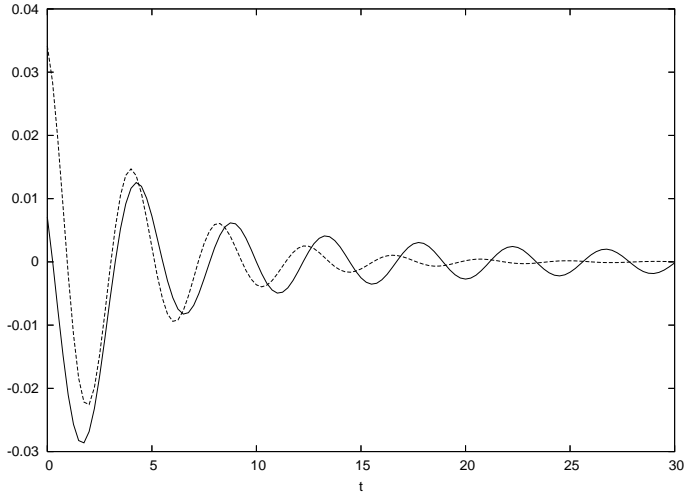}
\caption{$K^{(0)}(t)$ vs. $t$}
\label{fig2}
\end{figure}

\section{Results} 

\begin{figure}
\centering
\subfigure[$\bar{S}_x(t)$ vs. t]{
\includegraphics[scale=0.75]{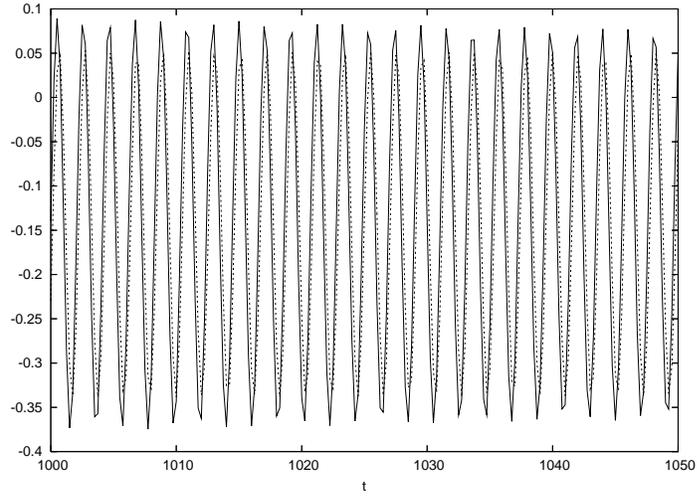}}
\subfigure[$\bar{S}_y(t)$ vs. t]{
\includegraphics[scale=0.75]{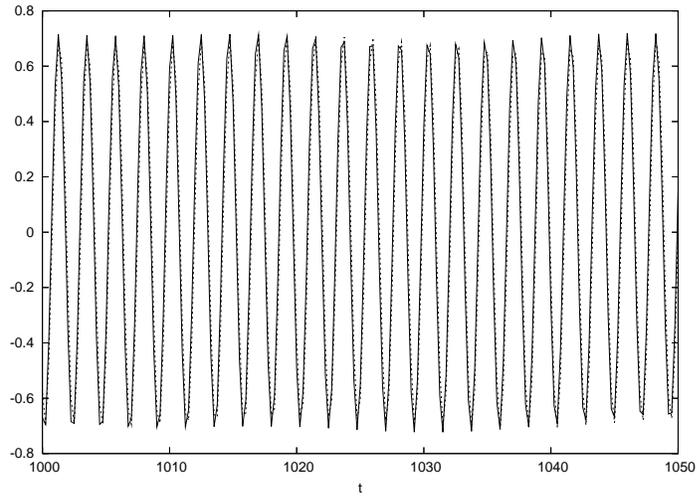}}
\subfigure[$\bar{S}_z(t)$ vs. t]{
\includegraphics[scale=0.75]{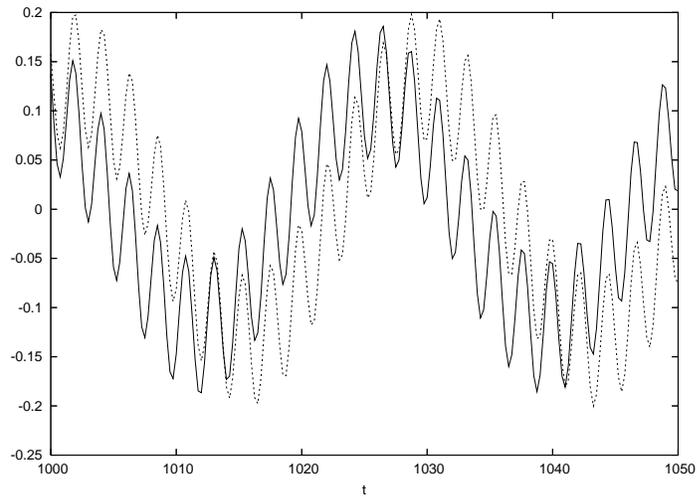}}
\caption{Mean spin matrices on $[1000,1050]$}
\label{fig3}
\end{figure}

\begin{figure}
\centering
\subfigure[$\bar{S}_x(t)$ vs. t]{
\includegraphics[scale=0.75]{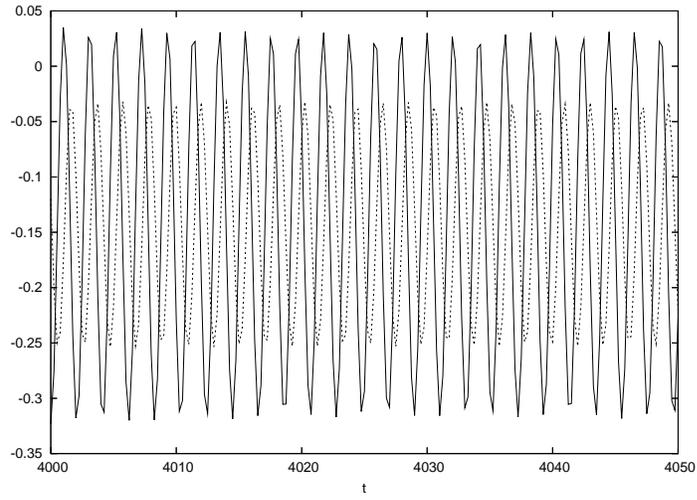}}
\subfigure[$\bar{S}_y(t)$ vs. t]{
\includegraphics[scale=0.75]{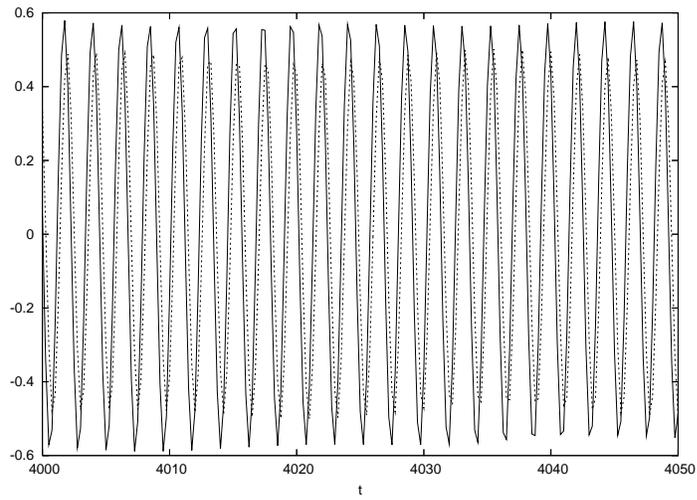}}
\subfigure[$\bar{S}_z(t)$ vs. t]{
\includegraphics[scale=0.75]{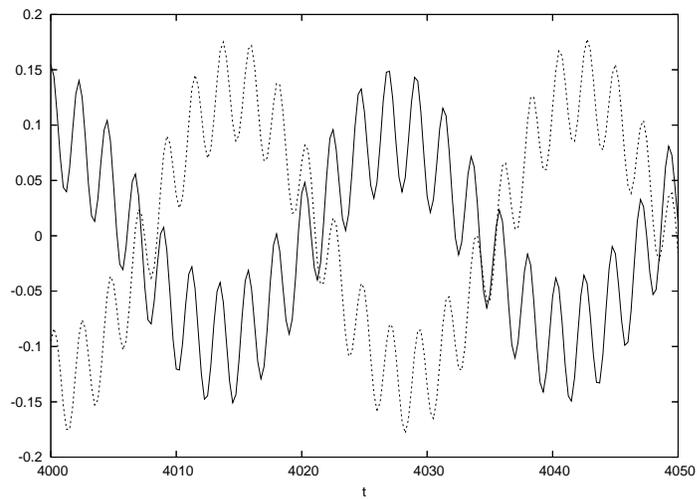}}
\caption{Mean spin matrices on $[4000,4050]$}
\label{fig4}
\end{figure}

\begin{figure}
\centering
\includegraphics[scale=0.75]{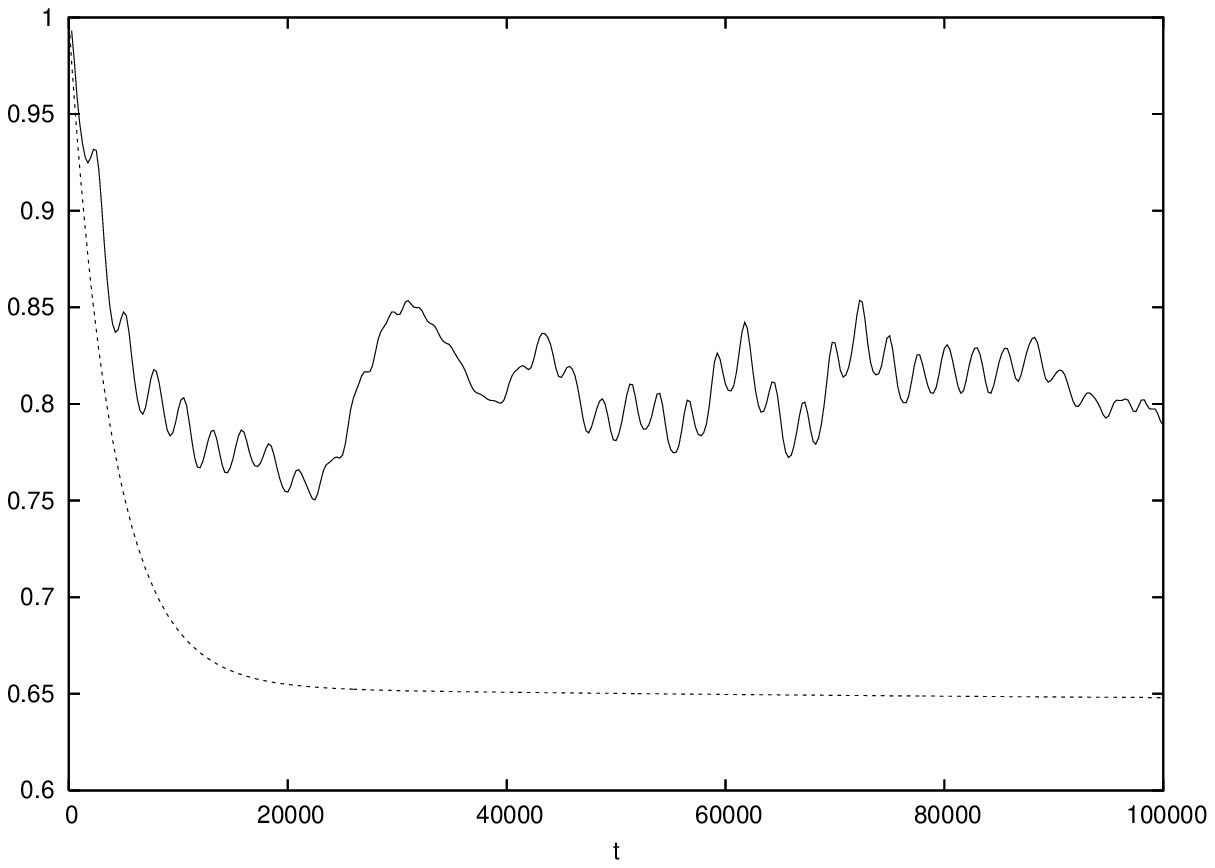}
\caption{${\cal P}(t)$ vs. $t$}
\label{fig5}
\end{figure}

The master equation and exact results are not visibly different until quite late in the dynamics.
Figure 3 shows the mean spin components $\bar{S}_x(t)={\rm Tr}\{S_X\rho(t)\}$, $\bar{S}_y(t)={\rm Tr}\{S_Y\rho(t)\}$, and $\bar{S}_z(t)={\rm Tr}\{S_Z\rho(t)\}$ computed exactly (solid curve)
and using the master equation (dashed curve) on the time interval $[1000,1050]$ (our time units are ns). By this point some discrepancy is discernible. The amplitudes of the oscillations
in $\bar{S}_x(t)$ are diminished compared to those in the exact calculations and they are slightly phase shifted. The results for $\bar{S}_y(t)$ are still in good agreement with the exact calculations. In the case of $\bar{S}_z(t)$ the error is larger but it appears mostly as a shift of the overall envelope of the oscillations. 

Figure 4 shows the mean spin components
on the time interval $[4000,4050]$. By this time the master equation results for $\bar{S}_x(t)$ and $\bar{S}_y(t)$ are further out of phase with their exact counterparts, and both amplitudes are diminished.
For $\bar{S}_z(t)$ the master equation results are completely out of phase with the exact results but the amplitudes are not any worse than they were at the earlier time.

Figure 5 shows the purity ${\cal P}(t)={\rm Tr}_S\{\rho(t)^2\}$ plotted on a much longer time interval $[0,100000]$ (or 0 to .1 ms). The solid curve is the exact result while the dashed is that of the SME. Here it is clear that there are serious problems with the predictions of the SME. A purity of about .8 should be seen at long times but the SME predicts just .65 which it out by nearly 20 \%.

To try to trace the origin of the errors we examined the diagonal elements of the reduced density operator in the eigenbasis of $H+\overline{{\cal B}}S$. These are shown in Figure 6 (a) for state 1 which is the lowest in energy, and for state 3 in (b) which is the highest in energy. The solid curves denote exact results while dashed curves indicate SME predictions. The relative errors in the lowest energy matrix element $\rho_{1,1}(t)$ are on the order of .1 \%. The errors in the highest energy matrix element $\rho_{3,3}(t)$ are on the order of 5 \%. The middle matrix element (not shown) is exact and constant because this eigenstate is also an eigenstate of $S_X$. Hence, the diagonal elements are not the primary source of the errors seen in Figure 5.

Figures 7-9 show the real and imaginary parts of $\sigma_{1,2}(t)$, $\sigma_{1,3}(t)$ and $\sigma_{2,3}(t)$ respectively in the eigenbasis of $H+\overline{{\cal B}}S$ where
$\sigma(t)=e^{i(H+\overline{{\cal B}}S)t}\rho(t)e^{-i(H+\overline{{\cal B}}S)t}$ is the reduced density matrix transformed to a rotating frame. Solid curves denote exact results while dashed curves are the SME predictions. Figure 7 for $\sigma_{1,2}(t)$ shows decent agreement between the SME and exact results until around 10000 time units. However, the master equation results decay to zero by about 40000 time units, while the exact results continue to oscillate with large amplitude. Similar results are seen in Figure 8 for $\sigma_{1,3}(t)$. Figure 9 for $\sigma_{2,3}(t)$ shows terrible agreement between the SME and exact results at all times except $t=0$.

The primary discrepancy between the master equation predictions and exact results thus arises in the off-diagonal elements in the long time limit. Moreover, the corresponding discrepancy in the purity suggests that these errors arise from problems with the dissipative terms in the SME. Since the results change little when we arbitrarily set $K^{(0)}(t)=0$ we also know that the inhomogeneous term is not the source of error.

\subsection{Numerical error}

We noticed an unexpected error arising in the eigenvalues of the density matrix computed using the SME master equation. One eigenvalue should remain zero at all times but we discovered that it becomes
slightly negative at short times but then returns to zero. This problem persists in double precision even for a requested absolute and relative tolerance of $10^{-16}$ but vanishes in quadruple precision with a requested absolute and relative tolerance of $10^{-17}$. Progamming in quad precision is impractical however since this greatly slows the computations. It does not seem likely that numerical errors of this type are responsible for the discrepancies between the exact and master equation results.

\begin{figure}
\centering
\subfigure[$\rho_{1,1}(t)$ vs. t]{
\includegraphics[scale=0.75]{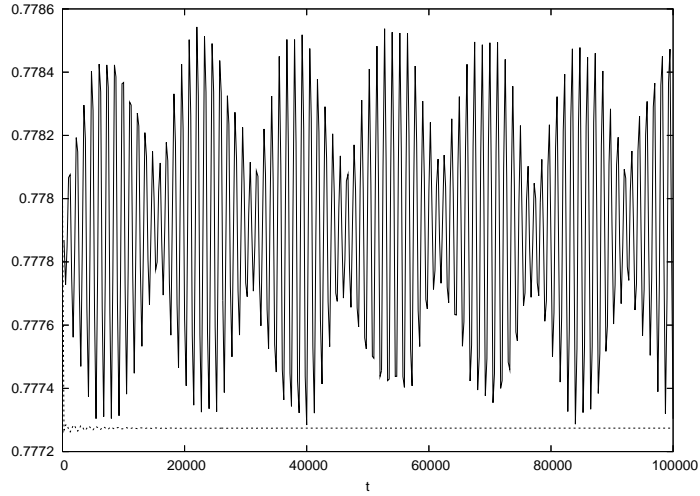}}
\subfigure[$\rho_{3,3}(t)$ vs. t]{
\includegraphics[scale=0.75]{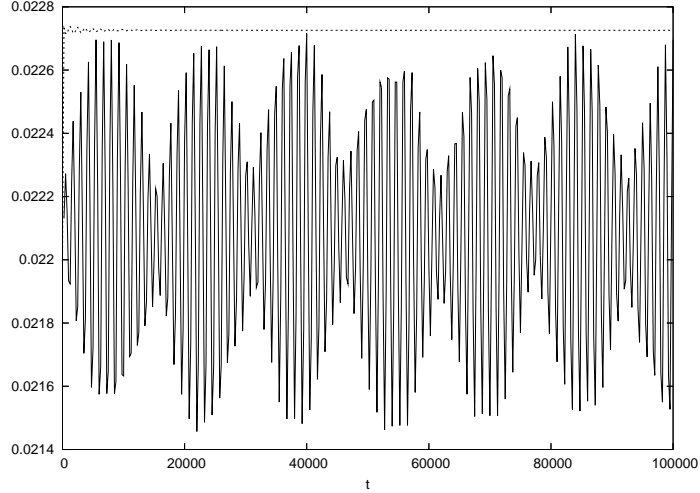}}
\caption{Diagonal elements of density matrix}
\label{fig6}
\end{figure}

\begin{figure}
\centering
\subfigure[${\rm Re}~\sigma_{1,2}(t)$ vs. t]{
\includegraphics[scale=0.75]{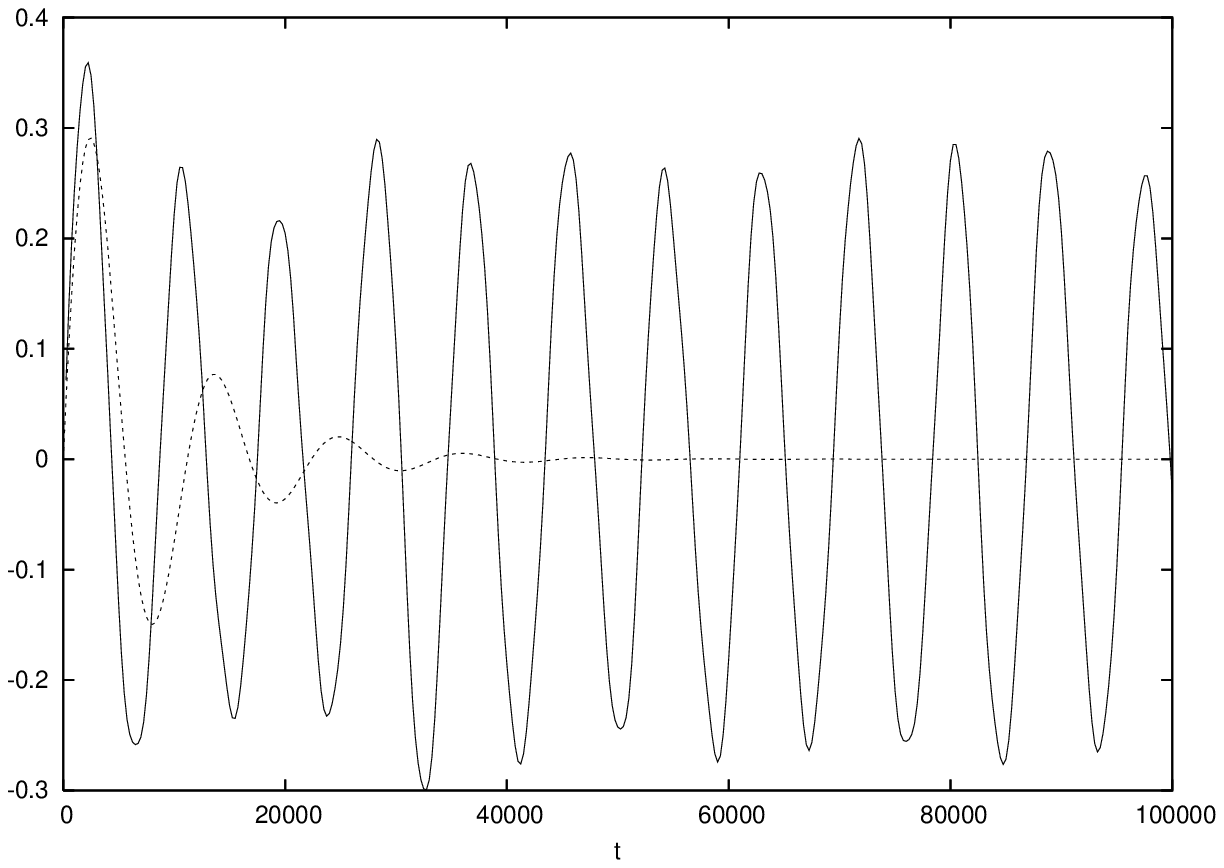}}
\subfigure[${\rm Im}~\sigma_{1,2}(t)$ vs. t]{
\includegraphics[scale=0.75]{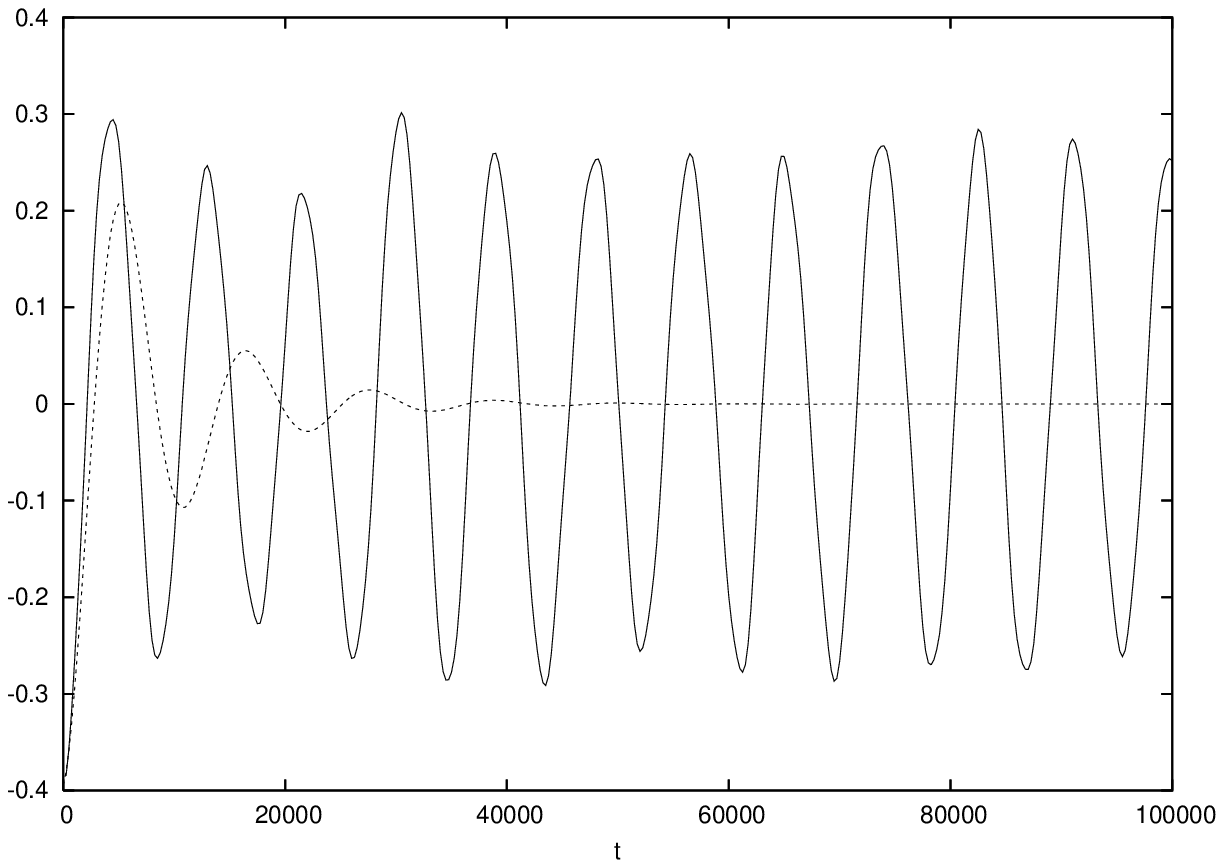}}
\caption{$\sigma_{1,2}(t)$ element of density matrix in rotating frame}
\label{fig7}
\end{figure}

\begin{figure}
\centering
\subfigure[${\rm Re}~\sigma_{1,3}(t)$ vs. t]{
\includegraphics[scale=0.75]{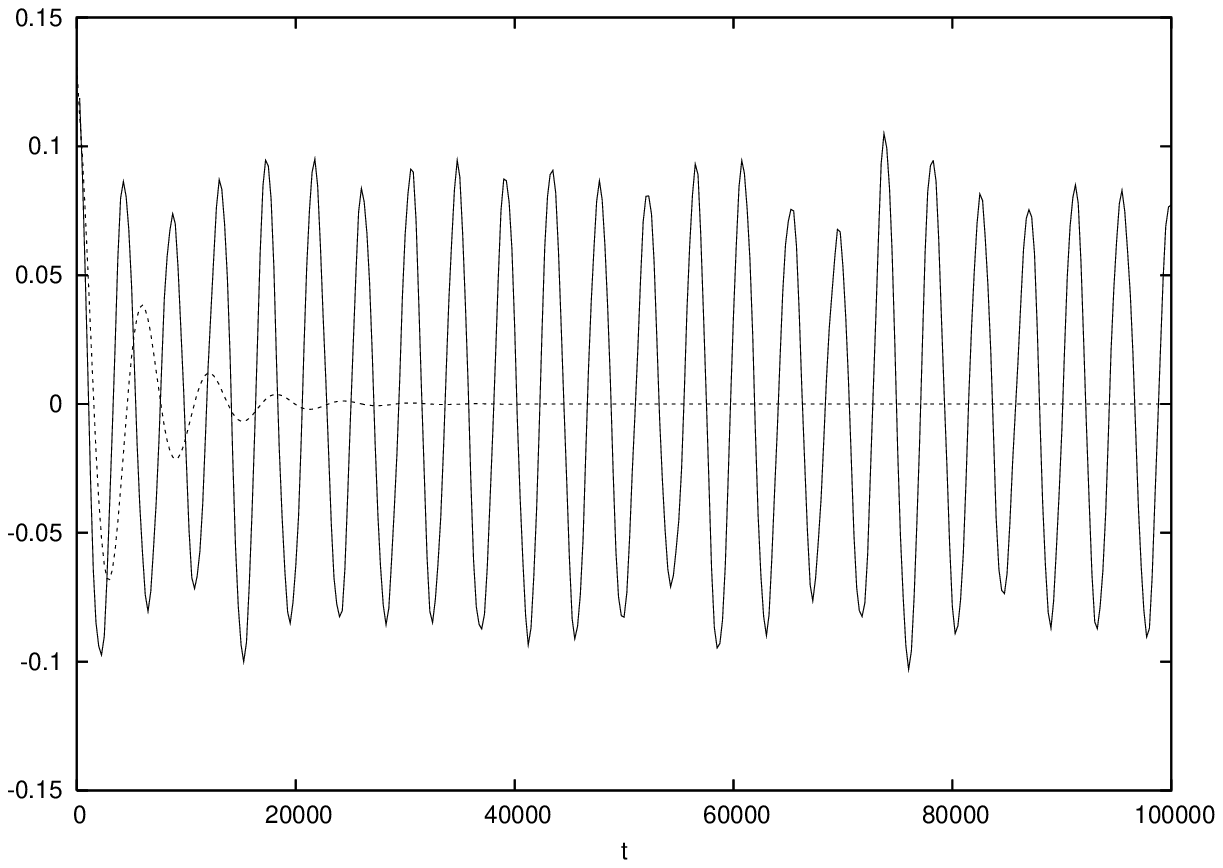}}
\subfigure[${\rm Im}~\sigma_{1,3}(t)$ vs. t]{
\includegraphics[scale=0.75]{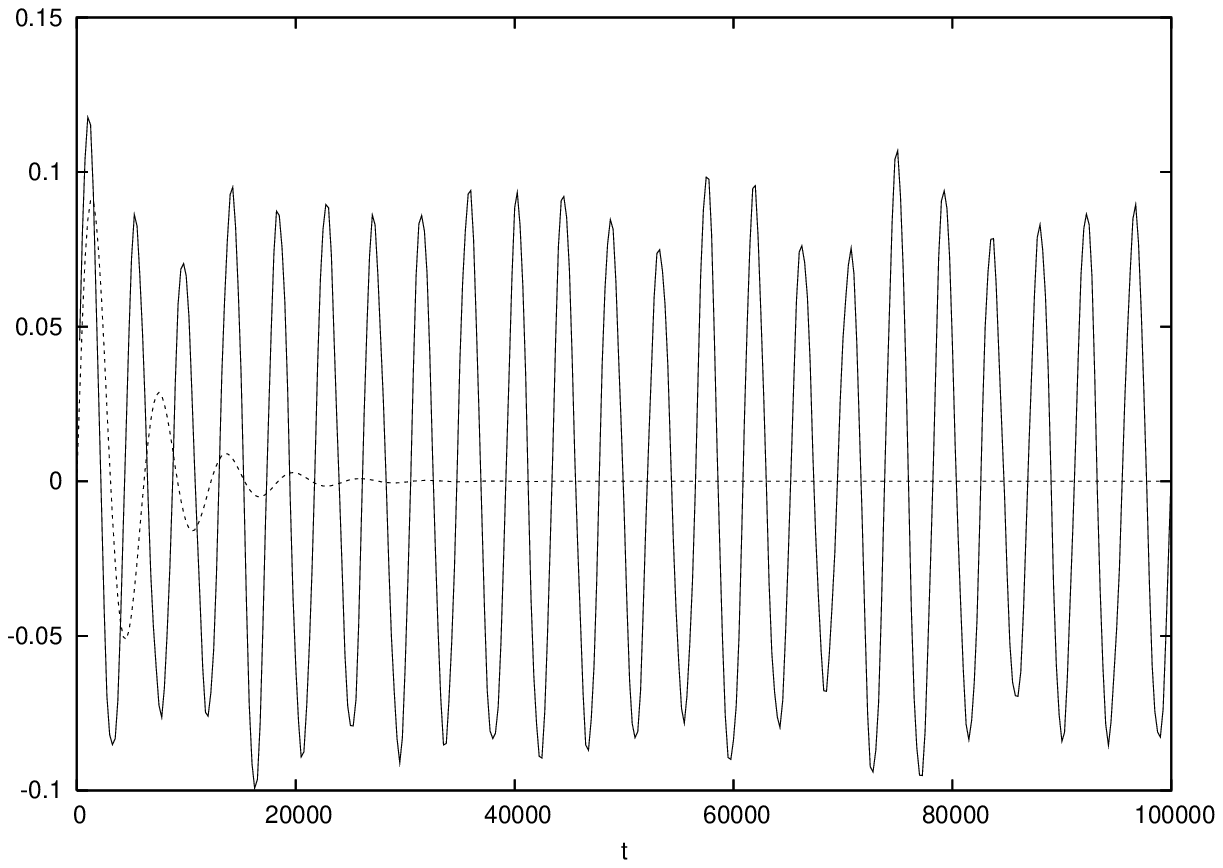}}
\caption{$\sigma_{1,3}(t)$ element of density matrix in rotating frame}
\label{fig8}
\end{figure}

\begin{figure}
\centering
\subfigure[${\rm Re}~\sigma_{2,3}(t)$ vs. t]{
\includegraphics[scale=0.75]{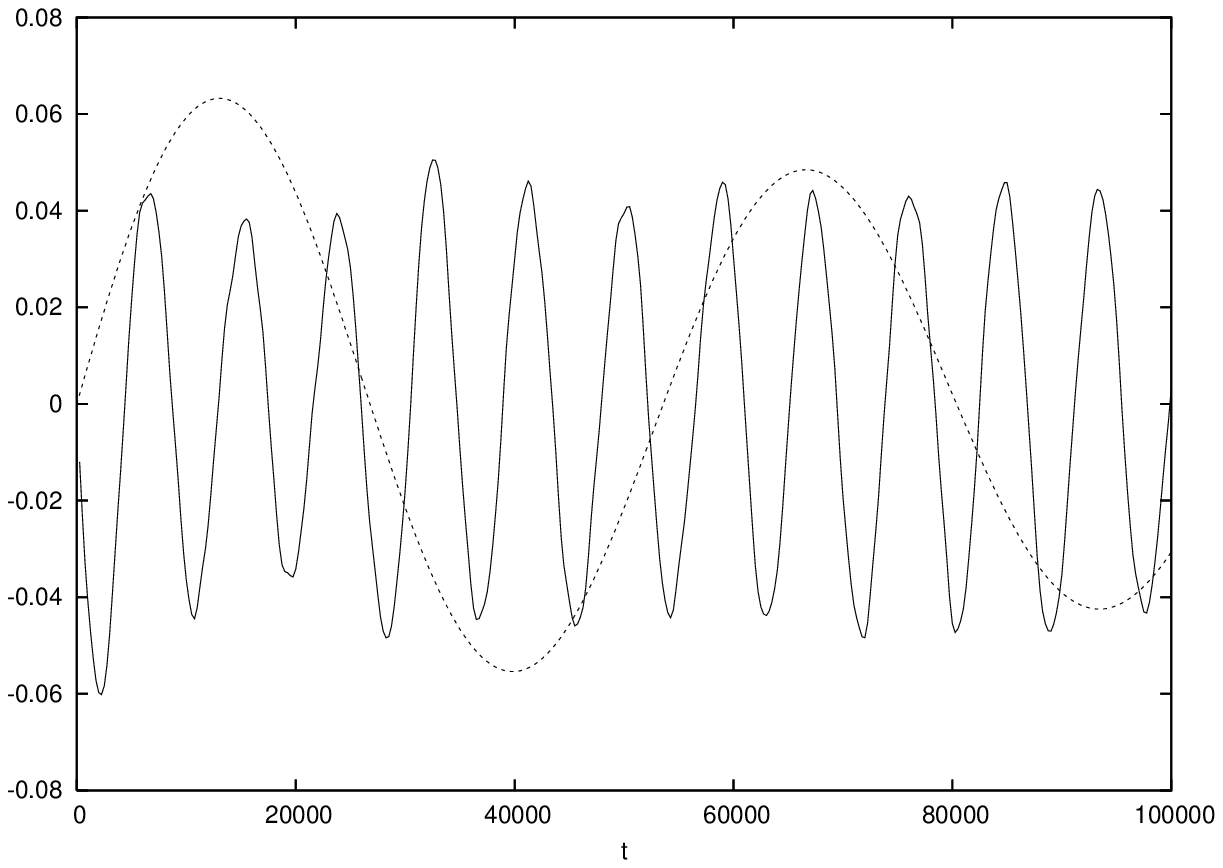}}
\subfigure[${\rm Im}~\sigma_{2,3}(t)$ vs. t]{
\includegraphics[scale=0.75]{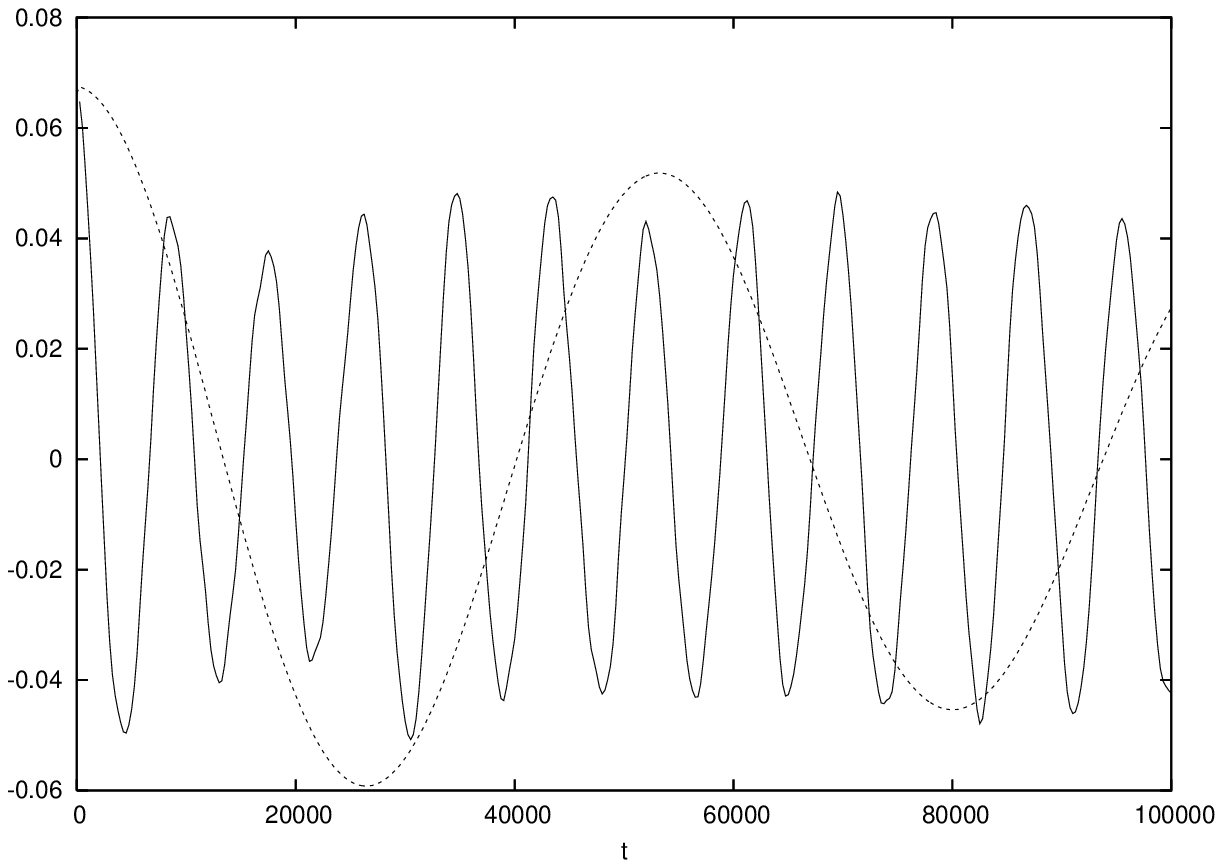}}
\caption{$\sigma_{2,3}(t)$ element of density matrix in rotating frame}
\label{fig9}
\end{figure}

\section{Discussion}

The synthetic approach introduced in this paper yields a well defined master equation with no free parameters which preserves positivity and correctly equilibrates to a long time limit which commutes with the shifted native system Hamiltonian and which depends at most on the diagonal elements of the initial density in the Hamiltonian eigenbasis. Numerical tests for a model system consisting of an NV center in diamond interacting with $^{13}$C impurities show good accuracy at short and intermediate times. The diagonal elements of the density in the shifted native Hamiltonian eigenbasis are also in good agreement with exact results at long times. There is however a serious problem with the off-diagonal elements. In a frame rotating with the native Hamiltonian the exact results show persistent oscillations with a period of about 10 $\mu$s, while the master equation results decay to zero quite rapidly. 

One possible solution to this problem would be to consider forms of the projection operator $\Lambda$ which do not commute with $H_B$ or $B$ in which case the mean field master equation would have a structure similar to Eq. (\ref{ME3}) but wherein the $K^{(2)}(t)$ memory function is non-zero. Clearly this structure will deviate from the standard  Kossakowski-Lindblad form but there may be some circumstances under which positivity is still preserved. It is also clear that this would change the off-diagonal elements but leave equilibration unaffected.

Alternatively, one might consider a more complex structure to accompany the modified $\Lambda$. Consider for example that one of the kernels of Eq. (\ref{ME2}) has the form
\begin{eqnarray}
\frac{1}{2}\left(M^{(3)}(t)-M^{(4)}(t)\right)=\frac{1}{2}{\rm tr}_B\{B[e^{-iQL_{tot}t},B]\Lambda\}
\end{eqnarray}
which can be rewritten using Kubo's formula\cite{Kubo} as
\begin{eqnarray}
&&\frac{1}{2}\left(M^{(3)}(t)-M^{(4)}(t)\right)=\nonumber \\
&&-\frac{i}{2}\int_0^tdt'~{\rm tr}_B\{B e^{-iQL_{tot}(t-t')} [QL_{tot},B] e^{-iQL_{tot}t'}\Lambda\}
\end{eqnarray}
which would suggest that $K^{(2)}(t)$ in Eq. (\ref{ME3}) be replaced by some operator like $K^{(3)}(t)[H,\cdot]+K^{(4)}(t) [S,\cdot]$ which would be of Kossakowski-Lindblad form if $K^{(3)}(t)=0$. Here $K^{(4)}(t)$ would then probably need to be proportional to $K^{(1)}(t)$ in order to preserve positivity.


\section*{References}

\begin{thebibliography}{99}

\bibitem{dsg} Kossakowski A 1972 {\it Rep. Math. Phys.} {\bf 3} 247; Lindblad G 1976 {\it Commun. Math. Phys.} {\bf 48} 119; \\
Gorini V, Kossakowski A, and Sudarshan E C G 1976 {\it J. Math. Phys.} 
{\bf 17} 821; \\
Alicki R and Lendi K 1987 {\it Quantum Dynamical Semigroups and
Applications} (Berlin: Springer)

\bibitem{BV} Breuer H P and Vacchini B 2009 \PR E {\bf 79} 041147

\bibitem{BG} Budini A A and Grigolini P 2009 \PR A {\bf 80} 022103

\bibitem{KR} Kossakowski A and Rebolledo R 2009 {\it Open Systems \& Information Dynamics} {\bf 16} 259

\bibitem{KR2} Kossakowski A and Rebolledo R 2008 {\it Open Systems \& Information Dynamics} {\bf 15} 1359

\bibitem{KR3} Kossakowski A and Rebolledo R 2007 {\it Open Systems \& Information Dynamics} {\bf 14} 265

\bibitem{Bud} Budini A A 2004 \PR A {\bf 69} 042107 

\bibitem{WW3} Wilkie J and Wong Y M 2011 \JPA {\bf 44} 275306

\bibitem{WW1} Wilkie J and Wong Y M 2009 \JPA {\bf 42} 015006

\bibitem{W1} Wilkie J 2000 \PR E {\bf 62} 8808

\bibitem{CK} Chru\'{s}ci\'{n}ski D and Kossakowski A 2010 {\it Phys. Rev. Lett.} {\bf 104} 070406

\bibitem{CKP} Chruscinski C, Kossakowski A and Pascazio S 2010 \PR A {\bf 81} 032101

\bibitem{BP2} Breuer H P and Piilo  J 2009 {\it Europhys. Lett.} {\bf 85} 50004

\bibitem{AB} Smirne A and Vacchini B 2010 \PR A {\bf 82} 022110

\bibitem{VB} Vacchini B and Breuer H-P 2010 \PR A {\bf 81} 042103

\bibitem{CKR} Chruscinski C, Kossakowski A and Rivas A 2011 \PR A {\bf 83} 052128

\bibitem{LPB} Laine E-M, Jyrki P and Breuer H-P 2010 \PR A {\bf 81} 062115

\bibitem{BLP} Breuer H-P, Laine E-M and Jyrki P 2009 \PRL {\bf 103} 210401

\bibitem{BP} See for example Breuer H P and Petruccione F 2002 {\it The Theory of Open Quantum Systems} (Oxford: Oxford University Press)

\bibitem{THERM1} Rajagopal A K 1998 \PL A {\bf 246} 237

\bibitem{THERM2} Dietz K 2003 \JPA {\bf 36} 5595

\bibitem{QC} Nielsen M A and Chuang I L 2000 {\it Quantum Computation and Quantum Information} (Cambridge University Press, Cambridge)

\bibitem{Wilk2} Wilkie J 2001 \JCP {\bf 114} 7736

\bibitem{Wilk} Wilkie J 2001 \JCP {\bf 115} 10335

\bibitem{Zwan} Nakajima S 1958 {\it Prog. Theor. Phys.} {\bf 20} 948; \\
Zwanzig R 1960 \JCP {\bf 33} 1338; \\
Zwanzig R 1961 in {\em Lectures in Theoretical Physics} Vol. 3 (New York: Interscience)

\bibitem{TW} Tessieri L and Wilkie J 2003 \JPA {\bf 36} 12305

\bibitem{RMT} Chalker J T and Wang Z J 2000 \PR E {\bf 61} 196; 1997 \PRL {\bf 79} 1797

\bibitem{RMT2} Haake F, Izrailev F, Lehmann N, Saher D and Sommers H-J 1992 {\it Z. Phys. } B {\bf 88} 359; Sommers H-J, Cristanti A, Sompolinsky H and Stein Y 1988 \PRL {\bf 60} 1895

\bibitem{T1} Linden N, Popescu S, Short A J and Winter A 2009 \PR E {\bf 79} 061103

\bibitem{T2} Lychkovskiy O 2010 \PR E {\bf 82} 011123

\bibitem{Fell} Feller W 1966 {\it An Introduction to Probability Theory and its Applications, Vol. 2} (New York: John Wiley and sons)

\bibitem{NKPPK} Nizovtsev A P, Kilin S Ya, Pushkarchuk V A, Pushkarchuk A L and Kuten S A 2010 {\it Optics and Spectroscopy} {\bf 108} 230

\bibitem{Dob} Hanson R, Dobrovitski V V, Feiguin A E, Gywat O and Awschalom D D 2008 {\it Science} {\bf 320} 352

\bibitem{Dob2} Dobrovitski V V, Feiguin A E, Hanson R and Awschalom D D 2009 \PRL {\bf 102} 237601

\bibitem{SIMANN} Software called {\it simann.f} available from NETLIB written by Goffe W L, Ferrier G D and Rogers J 1994 {\it Journal of Econometrics} {\bf 60} 65

\bibitem{Guru} Gurudev Dutt M V, Childress L, Jiang L, Togan E, Maze J, Jelezko F, Zibrov A S, Hemmer P R and Lukin M D 2007 {\it Science} {\bf 316} 1312

\bibitem{Note} In the absence of perturbations $E=0$ \cite{NKPPK}. Our value is set arbitrarily.

\bibitem{DOP853} dop853.f available from www.unige.ch/$\sim$hairer/prog/nonstiff/dop853.f

\bibitem{HW} Hairer E, Norsett S P and Wanner G 1993 {\it Solving ordinary differential equations I. Nonstiff problems, 2nd Ed.} (Springer-Verlag, Berlin)

\bibitem{Kubo} See Eq. 2.5 in Wilcox R M 1967 {\it J. Math. Phys.} {\bf 8} 962


\end{thebibliography}
\end{document}